\documentclass[twocolumn]{aastex631}
\usepackage{graphics,graphicx}
\hypersetup{urlcolor=blue}
\usepackage{natbib,hyperref}
\usepackage[outdir=./]{epstopdf}
\usepackage{textcomp,gensymb}
\usepackage{xfrac}
\usepackage{xcolor}
\usepackage{multirow}
\usepackage{longtable}

\newcommand{\mps}{m\,s$^{-1}$}

\newcommand{\vsini}{$v\sin{i_*}$}

\newcommand{\kepler}{{\it Kepler}}
\newcommand{\logg}{$log~g$ }

\newcommand{\um}{$\mu$m}
\newcommand{\fbol}{$F_{\mathrm{bol}}$}

\newcommand{\teff}{\ensuremath{T_{\text{eff}}}}
\newcommand\kms{km~s$^{-1}$}
\newcommand{\ms}{m~s$^{-1}$}

\newcommand{\gaia}{{\textit Gaia}}

\usepackage[outdir=./]{epstopdf}
\usepackage{ulem}
\usepackage{mathrsfs}
\usepackage{fontawesome}
\usepackage{comment}
\usepackage[version=4]{mhchem} 

\newcommand{\tess}{\textit{TESS}}

\newcommand{\jwst}{\textit{JWST}}

\newcommand{\starname}{TOI-6448}
\newcommand{\planetname}{TOI-6448\,b}

\shorttitle{A 34\,Myr giant planet} 
\shortauthors{Barber et. al.}

\begin{document}

\title{\tess\ Investigation - Demographics of Young Exoplanets (TI-DYE) IV: a Jovian radius planet orbiting a 34\,Myr Sun-like star in the Vela association}

\correspondingauthor{Madyson G. Barber}
\email{madysonb@live.unc.edu}

\author[0000-0002-8399-472X]{Madyson G. Barber}
\altaffiliation{NSF Graduate Research Fellow}
\affiliation{Department of Physics and Astronomy, The University of North Carolina at Chapel Hill, Chapel Hill, NC 27599, USA}

\author[0000-0003-3654-1602]{Andrew W. Mann}
\affiliation{Department of Physics and Astronomy, The University of North Carolina at Chapel Hill, Chapel Hill, NC 27599, USA}

\author[0000-0001-7246-5438]{Andrew Vanderburg}
\affiliation{Center for Astrophysics \textbar \ Harvard \& Smithsonian, 60 Garden Street, Cambridge, MA 02138, USA}
\affiliation{Department of Physics and Kavli Institute for Astrophysics and Space Research, Massachusetts Institute of Technology, Cambridge, MA 02139, USA}


\author[0000-0003-1464-9276]{Khalid Barkaoui}
\affiliation{Instituto de Astrof\'isica de Canarias (IAC), E-38200, La Laguna, Tenerife, Spain \label{IAC_Laguna}}
\affiliation{Astrobiology Research Unit, Universit\'e de Li\`ege, B-4000 Li\`ege, Belgium}
\affiliation{Department of Earth, Atmospheric and Planetary Science, Massachusetts Institute of Technology, Cambridge, MA 02139, USA}

\author[0000-0001-6588-9574]{Karen A.\ Collins}
\affiliation{Center for Astrophysics \textbar \ Harvard \& Smithsonian, 60 Garden Street, Cambridge, MA 02138, USA}

\author[0009-0006-9244-3707]{Sebastian Carrazco-Gaxiola}
\affiliation{Department of Physics and Astronomy, Georgia State University, Atlanta, GA 30302, USA}
\affiliation{RECONS Institute, Chambersburg, PA 17201, USA}

\author[0000-0002-5674-2404]{Phil Evans}
\affiliation{Phil Evans, El Sauce Observatory, Coquimbo Province, Chile}

\author[0000-0002-9641-3138]{Matthew J. Fields}
\affiliation{Department of Physics and Astronomy, The University of North Carolina at Chapel Hill, Chapel Hill, NC 27599, USA}

\author[0000-0003-1462-7739]{Michaël Gillon}
\affiliation{Astrobiology Research Unit, Universit\'e de Li\`ege, B-4000 Li\`ege, Belgium}

\author[0000-0002-9061-2865]{Todd J. Henry}
\affiliation{RECONS Institute, Chambersburg, PA 17201, USA}

\author[0000-0002-2135-9018]{Katharine~M.~Hesse}
\affiliation{Department of Physics and Kavli Institute for Astrophysics and Space Research, Massachusetts Institute of Technology, Cambridge, MA 02139, USA}

\author[0000-0003-0193-2187]{Wei-Chun Jao}
\affiliation{Department of Physics and Astronomy, Georgia State University, Atlanta, GA 30302, USA}

\author[0000-0001-8923-488X]{Emmanuel Jehin}
\affiliation{Space Sciences, Technologies and Astrophysics Research (STAR) Institute, Universit\'e de Li\`ege, B-4000 Li\`ege, Belgium}

\author[0000-0001-9827-1463]{Sydney Jenkins}
\affiliation{Department of Physics and Kavli Institute for Astrophysics and Space Research, Massachusetts Institute of Technology, Cambridge, MA 02139, USA}

\author[0009-0006-4398-4654]{Tim Johns}
\affiliation{Department of Physics and Astronomy, Georgia State University, Atlanta, GA 30302, USA}
\affiliation{RECONS Institute, Chambersburg, PA 17201, USA}

\author[0000-0003-1286-5231]{David~R.~Rodriguez}
\affiliation{Space Telescope Science Institute, 3700 San Martin Drive, Baltimore, MD, 21218, USA}

\author[0000-0001-8227-1020]{Richard P. Schwarz}
\affiliation{Center for Astrophysics \textbar \ Harvard \& Smithsonian, 60 Garden Street, Cambridge, MA 02138, USA}

\author[0009-0008-5045-1500]{William C. Storch}
\affiliation{Department of Physics and Astronomy, The University of North Carolina at Chapel Hill, Chapel Hill, NC 27599, USA}


\author[0000-0001-8621-6731]{Cristilyn N.\ Watkins}
\affiliation{Bozeman, MT 59718, USA}

\author[0000-0003-2127-8952]{Francis P. Wilkin}
\affiliation{Department of Physics and Astronomy, Union College, 807 Union St., Schenectady, NY 12308, USA}

\begin{abstract}
The discovery of infant ($<50$ Myr), close-in ($<$30-day period) planets is vital in understanding the formation mechanisms that lead to the distribution of mature transiting planets as discovered by \kepler. Despite several discoveries in this age bin, the sample is still too small for a robust statistical comparison to older planets. Here we report the validation of TOI-6448\,b, an $8.8\pm0.8\,R_\oplus$ planet on a 14.8 day orbit. TOI-6448 was previously identified to be a likely member of Vela Population IV. We confirm the star's membership and re-derive the age of the cluster using isochrones, variability, and gyrochronology. We find the star, and thus planet, to be $34\pm 3$\,Myr. Like other young planets, \planetname\ lands in a region of parameter space with few older planets. While just one data point, this fits with prior findings of an excess of 5-11$R_\oplus$ planets around young stars far beyond what can be explained by reduced sensitivity at young ages. Our ongoing search of Vela, Taurus-Auriga, Sco-Cen, and Orion are expected to reveal dozens more $<50$\,Myr transiting planets. 
\end{abstract}


\section{Introduction} \label{sec:intro}



Despite dozens of discoveries of planets $<$500 Myr, theories surrounding which formation mechanisms are dominant in reproducing the \kepler\ distribution are largely not understood. Two such pathways, the gas dwarfs versus water worlds formation, model sub-Neptunes and super-Earths as forming from the cores of young planets with once large, extended atmospheres \citep[e.g.][]{Lee2014, Ginzburg2016} or maintaining a consistent radii through time due to thinner, higher mean molecular weight atmospheres \citep[e.g.][]{Mordasini2009, Venturini2016, Zeng2019, Burn2024}. 

Testing these theories requires an understanding of the young planet population, as planets begin to lose markers of their formation quickly after the protoplanetary disk gas dissipates \citep[$\sim$\,3-10\,Myr; e.g.][]{Kenyon1995, Koepferl2013, Rogers2024}. Mass measurements, while able to distinguish between a true sub-Neptune or super-Earth \citep[e.g.][]{Wu2013, Wolfgang2016, Luque2022}, are difficult for young planets. Masses via radial velocities are complicated by stellar jitter \citep[e.g.][]{Blunt2023}, and masses through transit timing variations (TTVs) require detecting all major perturbing bodies \citep[small planets may elude detection due to young stellar noise or poor photometric precision, affecting the TTV analysis; e.g.][]{Weisserman2023} and rely on the system having multiple co-planar orbiting objects. Although transmission spectroscopy with \jwst\ has been successful in obtaining mass measurements of young planets \citep{deWit2013,Thao2024_featherweight, Barat2024}, obtaining enough mass measurements to test formation theories would require an intense observational campaign. These measurements can also be challenging for young sub-Neptunes, whose higher masses can mute spectral features. 

Instead, we can look at planet radii over time to approximate a typical evolution trajectory. For planet radii to be successful in differentiating the gas dwarf and water world scenarios, we rely heavily on understanding the youngest population ($<$50 Myr) where these formation pathways are most divergent \citep[e.g.][]{Rogers2025}. Even with multiple discovered systems in this age bin \citep[e.g. V\,1298\,$\tau$, HIP 67522, NGST 33, TIC 88785435;][]{David2019a_v1298, David2019b_v1298, THYMEII, Barber2024_hipc,  Alves2025_ngst33, Vach2025_tic887} and a distribution suggestive of a gas dwarf dominated formation scenario \citep{Vach2024_ocr}, we do not have \textit{enough} planets for a statistically significant differentiation between the two proposed pathways. This is further challenged by a number of these youngest planets orbiting fainter or more distant stars \citep[e.g. K2-33, TOI-1227, Kepler-1975, IRAS 04125+2902;][]{Mann2016_k233, THYMEVI_1227, Bouma2022, Barber2024_iras}, which can complicate follow up efforts.

A major goal of the \tess\ Investigation - Demographics of Young Exoplanets survey \citep[TI-DYE survey; see][]{Barber2024_hipc, Barber2024_iras} is discovering and characterizing the youngest transiting systems in order to test these formation pathways. The focus is on systems younger than 50\,Myr, where planet models make divergent predictions \citep[e.g.,][]{Karalis2025, Rogers2025} and before post-formation evolution erases information about a planet's initial conditions \citep{Marimbu2024}. A connected part of the survey is to find additional planets in known, young, transiting systems \citep[e.g.,][]{Barber2025_2076e} as multi-transiting systems are similarly powerful for testing models of planetary evolution \citep[e.g.][]{Dai2024}.

In this paper, we validate the detection of \planetname. We describe the observations of the system in Section \ref{sec:obs} and the planet detection in Section \ref{sec:detection}. In Section \ref{sec:stellarprops} and Section \ref{sec:planetprops}, we derive the star and planet properties, respectively, and perform an injection-recovery analysis in Section \ref{sec:injrec}.
We discuss false positive scenarios in Section \ref{sec:fp}. We use the list of comoving stars identified in Section \ref{sec:friendfinder} to derive a precise age for the system in Section \ref{sec:age}. Finally, we discuss the importance of this system and continued work in Section \ref{sec:concl}.

\section{Observations} \label{sec:obs}

\subsection{\tess\ Light Curve}\label{sec:tessobs}

TOI 6448 (TIC 320411045) was first observed by \tess\ in Sector 33 from 2020 December 18 to 2021 January 13, and re-observed in sectors 34 (2021 January 14 - 2021 February 8), 61 (2023 January 18 - 2023 February 12), and 87 (2024 December 18 - 2025 January 14). The target was pre-selected for 2-minute and 20-second short cadence light curves for Sector 87, and in the remaining sectors, the target was only observed in the \tess\ full frame images (FFIs). The \tess\ data used in this analysis can be found in MAST \citep{MAST_SPOC_LCs, MAST_FFI_LCs}.

We built our \tess{} light curve following \citet{Barber2024_iras}, which was an update of the extraction and systematic corrections of \citet{Vanderburg2019}\footnote{we call these VanderCurves}. This method has been used effectively on young variable stars with transiting planets \citep[e.g.,][]{Capistrant2024, Barber2024_hipc, Thao2024_1224}. We always use the light curve with the fastest cadence available for each sector. 

\subsubsection{Planet Detection}\label{sec:detection}

The QLP (Quick-Look Pipeline) faint-stars search \citep{Kunimoto2022_QLPfaintstar} of sectors 33, 34, and 61 identified a 14.8 day signal, which was released as \starname.01 on 2023 May 3.

We ran the updated \texttt{Notch \& LOCoR} \citep[\texttt{N\&L};][]{Rizzuto2017} as described in \cite{Barber2024_iras}. Using a 0.5 day filtering window, we used \texttt{Notch} to detrend the light curve with a second-order polynomial while preserving trapezoidal, transit-like shapes. As it detrends the light curve, for every point, \texttt{Notch} calculates the change in the Bayesian Information Criterion (BIC) based on how well adding the trapezoid to the polynomial improved the model. We then implemented a box-least squares (BLS) search on the BIC time-series to search for periodic signals between 0.5 and 30 days with an SNR $>$8. We recovered the 14.8 day signal with a BLS SNR of 33. The detected period, initial transit time, and depth matched that of the \tess\ QLP detection. No other planet-like signals were detected with periods 0.5-30\,days.

\subsection{Ground-based Photometry}

The \textit{TESS} pixel scale is $\sim 21\arcsec$ pixel$^{-1}$ and photometric apertures typically extend out to roughly 1\arcmin, generally causing multiple stars to blend in the \textit{TESS} photometric aperture. Multi-wavelength observations of the transit are also useful for false-positive vetting. To these ends, we acquired ground-based time-series follow-up photometry of the field around TOI-6448 as part of the \textit{TESS} Follow-up Observing Program \citep[TFOP;][]{collins:2019}\footnote{\url{https://tess.mit.edu/followup}}.

\subsubsection{LCOGT\label{subsec:lcogt}}

We observed an ingress window and a full transit window of TOI-6448.01 on UTC 2024 November 25 and UTC 2024 December 09, respectively, in Sloan $g'$ band from Las Cumbres Observatory Global Telescope \citep[LCOGT;][]{Brown:2013} 0.35\.m network nodes. The November light curve was observed from Teide Observatory on the island of Tenerife (TEID), and the December light curve was observed from South Africa Astronomical Observatory near Sutherland, South Africa (SAAO). The 0.35\,m Planewave Delta Rho 350 telescopes are equipped with a $9576\times6388$ QHY600 CMOS camera which has an image scale of $0.73$\arcsec\ per pixel, resulting in a $114\arcmin\times72\arcmin$ full field of view. We used the optional $30\arcmin\times30\arcmin$ sub field of view for a faster detector read-out. 

The images were calibrated by the standard LCOGT {\tt BANZAI} pipeline \citep{McCully:2018} and differential photometric data were extracted using {\tt AstroImageJ} \citep{Collins:2017}. We used circular $8.7$\arcsec\ and $3.7$\arcsec\ photometric apertures for the November and December observations, respectively.
We detected the transit in the target star photometric aperture in both light curves, which confirms that the \tess\ detected event is indeed occurring in TOI-6448. Furthermore, the Sloan $g'$ band transit depth is consistent with the transit depths in the redder \tess\ band and other follow-up bands (see Section~\ref{sec:planetprops}).

\subsubsection{TRAPPIST-South}
We used the TRAPPIST-South \citep[TRAnsiting Planets and PlanetesImals Small Telescope,][]{Jehin2011,Gillon2011} telescope to observe a full transit of TOI-6448.01 on UT 2024 February 17.
TRAPPIST-South is a 0.6m Ritchey-Chr\'etien telescope located at La Silla Observatory in Chile, and it is equipped with a 2K$\times$2K FLI Proline CCD camera with a pixel scale of 0.65\arcsec,\, and a $22\arcmin\times22\arcmin$ field of view. 
The observations were conducted in the $I+z$ filter with an exposure time of 50s. During the observations of the target, the telescope underwent a meridian flip at BJD 2460357.601.

Data reduction and photometric extraction were performed using the {\tt AstroImageJ}  \citep{Collins:2017} software using a 9-pixel (6.6\arcsec) aperture radius. As with LCO, the transit was clearly visible and exhibited a depth consistent with the \tess{} transits. 

\subsubsection{El Sauce}
We observed a full transit in Johnson-Cousins $R_c$-band on UT 2024 February 17 using the Evans 0.51m telescope at El Sauce Observatory in Coquimbo Province, Chile. The telescope was equipped with a Moravian C3-26000 camera with resolution 6252 $\times$ 4176 pixels. After binning 2 $\times$ 2 in camera the resulting image scale was 0.449\arcsec\ per pixel. The photometric data was obtained from 150 $\times$ 120 seconds exposures, after standard calibration, using a circular 5.4\arcsec\ aperture in {\tt AstroImageJ} \citep{Collins:2017}. The transit depth was consistent with the \tess\ transits.

\subsection{Ground-based Spectroscopy}

\begin{table}[]
    \centering
    \begin{tabular}{lccr}
    \hline
    \hline
    Epoch & RV & $\sigma_{RV}$ & Instrument\\
    (BJD) & (\kms) & (\kms) & \\
    \hline
    2460819.467 & 21.40 & 1.05 & CHIRON \\
    2460822.459 & 21.34 & 2.03 & MIKE \\
    \hline
    -- & 19.10 & 2.06 & \gaia\ DR3 \\
    \hline
    \end{tabular}
    \caption{Radial velocity measurements of \starname\ taken in this work. The \gaia\ RV is listed for reference.}
    \label{tab:rvs}
\end{table}

\subsubsection{CHIRON}\label{sec:chiron}
We observed \starname\ on UT 2025 May 23 using the CHIRON echelle spectrometer on the SMARTS 1.5 m telescope at the Cerro Tololo Inter-American Observatory in Chile. The exposure was 20 minutes and utilized the fiber mode. CHIRON fiber mode covers 4100-8700 $\AA$ with a resolution R$\sim$25000. We reduced the data using the CHIRON data reduction pipeline \citep{Tokovinin2013, Paredes2021}.

We extract the radial velocity by cross-correlating the continuum-normalized spectra to PHOENIX model spectra \citep[5900K, $log\,g = 4.5$;][]{Husser2013} and correcting for barycentric motion using \texttt{barycorrpy} \citep{Wright2014_barycorrpy}. For RVs, we used orders with strong absorption features (e.g. CaII, Li, H, He, and K) and adopted the median value and standard deviation as the RV for this epoch (Table \ref{tab:rvs}). 

\subsubsection{MIKE}\label{sec:mike}

We observed \starname\ on UT 2025 May 26 using the Magellan Inamori Kyocera Echelle \citep[MIKE;][]{Bernstein2003} instrument on the Magellan Clay telescope at the Las Campanas Observatory in Chile using the 1.0" slit and both arms (red (R$\sim$22,000) and blue (R$\sim$28,000)) simultaneously, covering 3350-9500 $\AA$. We reduced the 15 minute exposure using the CarPy MIKE pipeline \citep{Kelson2000, Kelson2003}.

Repeating the same processes as for the CHIRON spectra, we cross-correlate the continuum-normalized spectra to PHOENIX model spectra, using only orders with strong stellar absorption features, and correcting for barycentric motion. We again adopt the median value and standard deviation (across both arms) as the reported value for this epoch.

\section{Stellar Properties}  
\label{sec:stellarprops}

\begin{table}[]
    \centering
    \caption{Stellar Parameters of TOI-6448}
    \begin{tabular}{lcc}
    \hline
    \hline
    Parameter & Value & Source \\
    \hline
    \multicolumn{3}{c}{Identifiers}\\
    \hline
    TOI & 6448 & \tess\ \\
    TIC & 320411045 & \tess\ \\
    Gaia & 5590921485126584832 & \gaia\ DR3  \\
    \hline
    \multicolumn{3}{c}{Astrometry}\\
    \hline
    $\alpha$ & 110.170410 & \gaia\ DR3 \\
    $\delta$ & -33.922328 & \gaia\ DR3  \\
    $\mu_\alpha$ (mas yr$^{-1}$) & $-8.021 \pm 0.010$ & \gaia\ DR3 \\
    $\mu_\delta$ (mas yr$^{-1}$) & $5.0780 \pm 0.0128$ & \gaia\ DR3 \\
    $\pi$ (mas)  & $2.637 \pm 0.011$ & \gaia\ DR3 \\
    \hline
    \multicolumn{3}{c}{Photometry}\\
    \hline
    \tess\ (mag)  & $12.3239 \pm 0.0061$ & \tess\ \\
    $G$ (mag) & $12.7708\pm0.0029$ & \gaia\ DR3 \\
    $B_P$ (mag) & $13.1151\pm0.0040$ & \gaia\ DR3 \\
    $R_P$ (mag) & $12.2579\pm0.0043$ & \gaia\ DR3 \\
    \hline
    \multicolumn{3}{c}{Physical Properties}\\
    \hline
    \vsini (\kms) & $20.5\pm1.7$ & This work \\
    $i_*$($\degree$) & $>70$ & This work \\
    $P_{rot}$ (days) & $2.412\pm0.037$ & This work \\
    $F_{bol}$ (erg cm$^{-2}$ s$^{-1}$) & $(2.00\pm0.12)\times10^{-10}$ & This work \\
    $T_{eff}$ (K) & $5910\pm90$ & This work \\
    $A_V$ (mag) & $0.06^{+0.06}_{-0.04}$ & This work \\
    $R_*$ ($R_\odot$) & $ 0.897\pm0.07$ & This work \\
    $M_*$ ($M_\odot$) & $1.03\pm0.05$ & This work \\
    $\rho_*$ ($\rho_\odot$) & $1.43\pm0.37$ & This work \\
    $L_*$ ($L_\odot$) & $0.908\pm0.068$ & This work \\
    Age (Myr) & $34\pm 3$ & This work \\
    \hline
    \end{tabular}
    \label{tab:starparams}
\end{table}

\subsection{\teff, $L_{*}$ and $R_{*}$ from the SED}\label{sec:sed}

\begin{figure}
    \centering
    \includegraphics[trim={0 0 80 430},clip,width=0.48\textwidth]{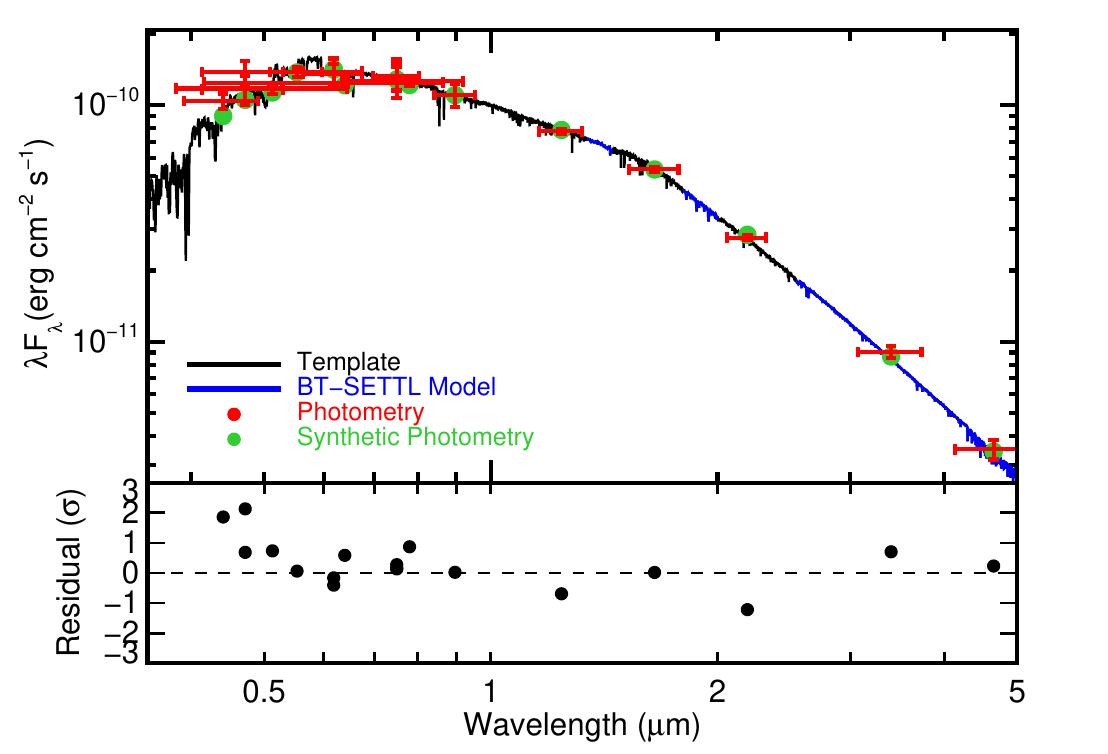}
    \caption{Fit to the SED of \starname{}, showing the best-fit template (black) filler model (blue), observed photometry (red) and synthetic photometry (green). For the photometry, the vertical error bars are the photometric uncertainties, while the horizontal error bars represent the filter width. The bottom panel shows the residuals in units of standard deviations. The fit is marginalized over a range of templates and extinctions; this one is a G1.5V with low extinction ($A_V=0.05$).}
    \label{fig:sed}
\end{figure}

We fit the spectral-energy distribution (SED) of \starname\ following \citet{Mann2016_k233}. Specifically, we gathered photometry from \citet{Skrutskie2006}, \citet{Henden2012}, \citet{allwise}, \citet{Skymapper1}, or \citet{Evans2018}. We compared the photometry to a grid of flux-calibrated templates from \citet{Heap2007}, \citet{Rayner2009}, and \citet{Villaume2017}, which have been supplemented with optical or NIR flux-calibrated spectra where available \citep{Mann2013c,Gaidos2014}. The templates generally span 0.36--2.4\um, but we used PHOENIX BT-SETTL atmosphere models \citep{Allard2013} to fill from 2.4--20\um\ and a few regions of high telluric contamination. 

To compute \fbol, we integrated the resulting absolutely-calibrated spectrum, and turned this into $L_*$ using the \gaia\ DR3 parallax. We estimated \teff\ from the BT-SETTL model fit and $R_*$ from the Stefan-Boltzmann relation as well as the scaling of the model to the data \citep[IRFM;][]{Blackwell1977}. 

As we show in Figure~\ref{fig:sed}, the overall fit to the photometry was excellent; the best-fit had a $\chi^2_\nu\lesssim1$ with no sign of systematics with wavelength. The resulting parameters were \fbol$=(2.00\pm0.12)\times10^{-10}$erg\,cm$^{-2}\,s^{-1}$, $L_*=0.908\pm0.068L_\odot$, \teff=$5910\pm90$\,K and $R_*=0.897\pm0.070R_\odot$. We found extinction to be low ($A_V<0.15$), with some templates reproducing the photometry without any extinction corrections. Final parameters are reported in Table~\ref{tab:starparams}

\subsection{$M_*$ from $T_{eff}$ and Age}

We estimate the stellar mass following the procedure in \cite{Fields2025}. To summarize, \texttt{stelpar}\footnote{\url{https://github.com/mjfields/stelpar}} simultaneously fits for stellar mass ($M_*$), age, extinction ($A_V$), and underestimated uncertainties ($f$) by comparing observed photometry to evolutionary model grids in an MCMC framework. We ran \texttt{stelpar} against the DSEP-magnetic \citep{Feiden2016} and PARSEC \citep{Bressan2012} models, placing a prior on $T_{eff}$ ($T_{eff}$ = $5910 \pm 90$ K) and a loose prior on age ($30 \pm 10$ Myr; see Section \ref{sec:age}). The model also produces a stellar radius, which we can compare to our SED fit as a check. The DSEP-magnetic model produced $M_* = 1.12\pm0.03 M_\odot$, $A_V = 0.46\pm0.2$, and $R_* = 1.09\pm0.03 R_\odot$, while the PARSEC model produced $M_* = 1.03\pm0.01 M_\odot$, $A_V = 0.06^{+0.06}_{-0.04}$ and $R_* = 0.92\pm0.01 R_\odot$. Due to the stellar radius and extinction agreeing with the SED fit, we choose to adopt the PARSEC model for our stellar mass. 

Based on a comparison to young stars with transit-based densities, \citet{Fields2025} recommends maintaining a minimum uncertainty of 5\% on the produced values \citep[also see][]{Tayar2022}, so we report a final stellar mass of $M_* = 1.03\pm0.05 M_\odot$. For $R_*$, we use the value from the SED fit (Section~\ref{sec:sed}), which we note agrees well with the PARSEC radius ($<1\sigma$) and marginally with the DESP-mag radius (2.5$\sigma$).

\subsection{Rotational broadening and Stellar inclination}

We estimate the rotational broadening (\vsini) of \starname\ by cross-correlating the observed MIKE and CHIRON spectra with the PHOENIX model spectra. We adapted the procedure (and associated code\footnote{\url{https://github.com/aurorayk/Vsini}}) from \citet{Kesseli2018}, with minor updates described in \citet{Fields2025}. To summarize, we artificially broadened the slow-rotating PHOENIX model spectra using a grid of \vsini\ values linearly spaced from 2--50\kms\ and cross-correlated each with the observed spectra. We then measure the full-width-half-maximum of the cross-correlation function to determine the \vsini. We repeat this process for each instrument, using the same orders used to find the radial velocities in Section \ref{sec:chiron} and \ref{sec:mike}. We drop three values which disagreed with the remaining 12 orders (likely due to stronger tellurics and weaker stellar lines).

Taking the median and standard deviation of the 12 \vsini\ measurements, we determine \vsini $ = 20.5\pm1.7$\,\kms. We find that perturbing the model metallicity (-0.5, 0.5), \logg (4.0, 5.0), and temperature (5800, 6000 K) did not significantly alter the \vsini\ measurement outside of uncertainties.

Following \citet{Masuda2020} and using the associated code from \cite{Fields2025}\footnote{\url{https://github.com/mjfields/cosi}}, we estimate the stellar inclination ($i_*$) from \vsini. Taking into account the uncertainties in the rotation period (calculated with the rest of the identified co-moving stars in Section \ref{sec:gyro}), stellar radius, and \vsini, we find $i_* > 70\degree$ at $1\sigma$ ($i_* > 62\degree$ at $2\sigma$). This is consistent with edge-on stellar rotation, as we expect if the planet is aligned with the host.

\section{Planet Properties} 
\label{sec:planetprops}

\subsection{\tess\ Transit Fit}\label{sec:tessFit}

\begin{figure*}
    \includegraphics[width=0.33\textwidth]{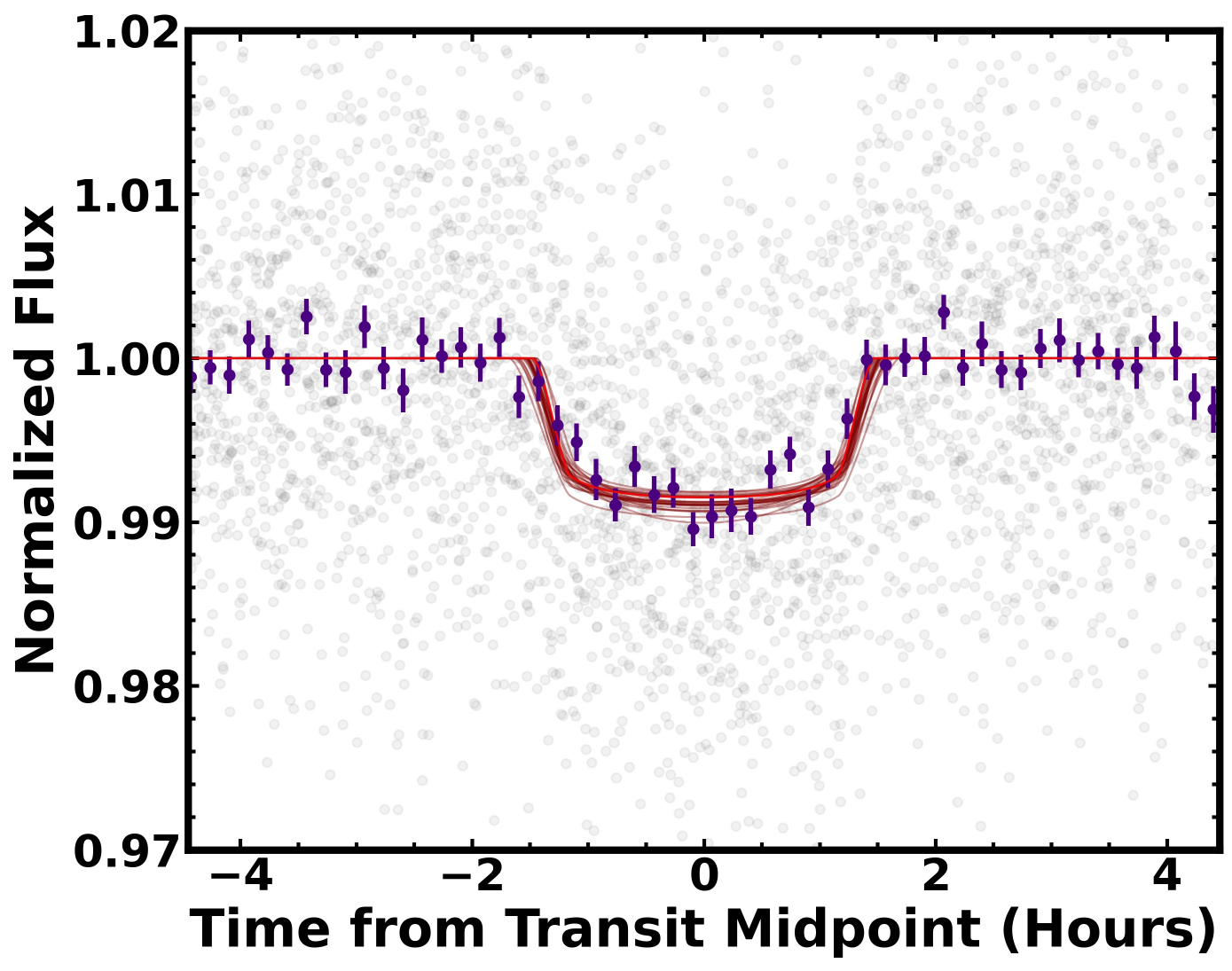}
    \includegraphics[width = 0.665\textwidth]{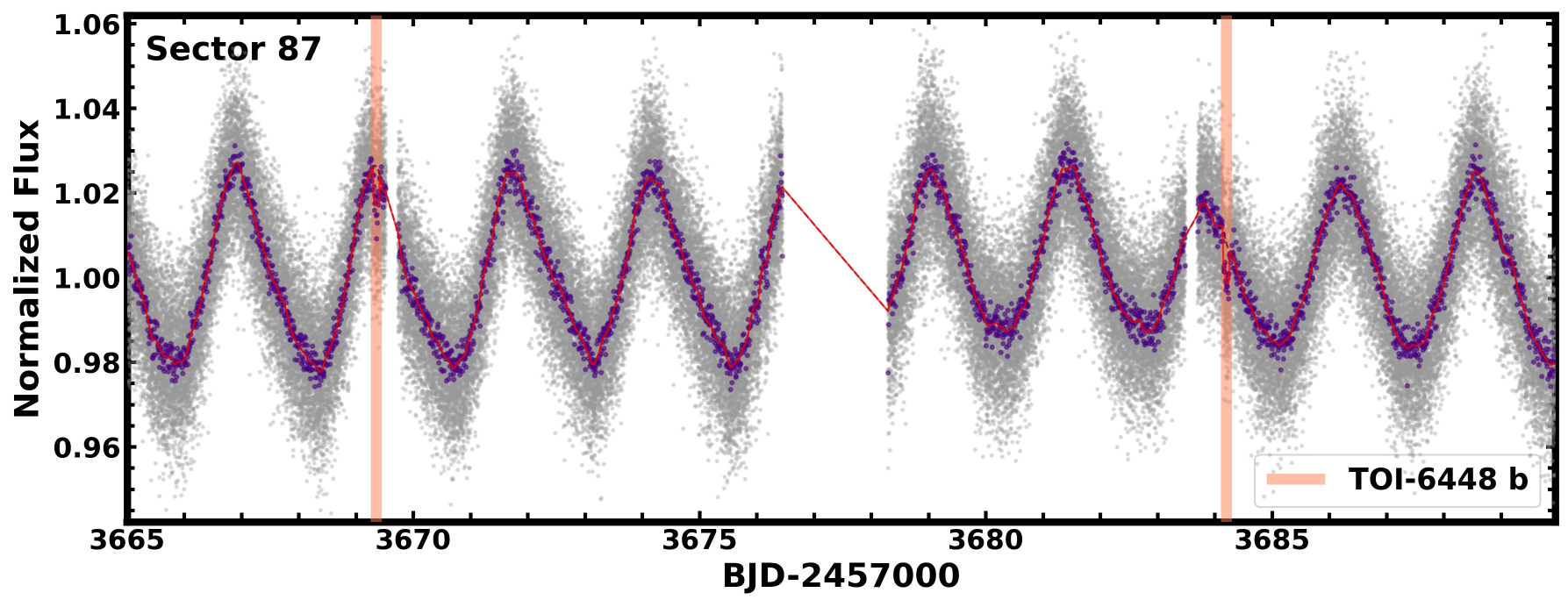}
    \caption{Left) Phase-folded \tess\ light curve of TOI-6448 (gray) binned to 10-minute intervals (purple) for clarity. The best-fit transit model is shown as the bright, opaque red line with 25 sample fits pulled from the posterior shown as the dark, translucent red lines. The GP has been removed from the data and model. Right) Representative section of the \tess\ light curve (gray) binned to 10-minute intervals (purple) for clarity. The red line shows the GP model, and the pink lines show the locations of the transit events. }
    \label{fig:lc}
\end{figure*}

\begin{table}[]
    \centering
    \caption{Priors used in the \texttt{MISTTBORN} \tess\ transit fit.}
    \begin{tabular}{lcc}
    \hline
    \hline
     Description  &  Parameter & Prior$^\alpha$ \\
     \hline
     \multicolumn{3}{c}{Planet Parameters}\\
     \hline
     impact parameter & $b$ & $U(0.0,1.0)$ \\
     planet-to-star radius ratio &  $R_P$/$R_*$ & $U(0.0,1.0)$ \\
     \hline
     \multicolumn{3}{c}{Stellar Parameters}\\
     \hline
     limb-darkening coefficient & $q1$ & $U(0.0,1.0)$ \\
     limb-darkening coefficient & $q2$ & $U(0.0,1.0)$ \\
     stellar density & $\rho_*$ ($\rho_\odot$) & $N(1.42,0.37)$ \\
     \hline
     \multicolumn{3}{c}{GP Parameters}\\
     \hline
     period & $P$ (days) & $P>0$ \\
     damping timescale & $\tau$ (days) & $\tau>P$\\
     standard deviation & $\sigma$ & $\sigma > 0$\\
     \hline
     \multicolumn{3}{p{\linewidth}}{$\alpha$ $U(a,b)$ indicates a uniform distribution from $a$ to $b$.\newline\hspace*{2.75mm}$N(a,b)$ indicates a normal distribution centered at $a$\newline\hspace*{2.75mm}with a standard deviation of $b$.}
    \end{tabular}
    \label{tab:priors}
\end{table}

We fit the systematic-corrected \tess\ light curves using \texttt{MISTTBORN} \citep[MCMC Interface for Synthesis of Transits, Tomography, Binaries, and Others of Relevant Nature;][]{Mann2016a, MISTTBORN}\footnote{\url{https://github.com/captain-exoplanet/misttborn}}. \texttt{MISTTBORN} simultaneously fits for the planet and stellar variability using \texttt{BATMAN} transit models \citep{BATMAN} and a \texttt{celerite2} Gaussian Process \citep[GP;][]{celerite2} in a MCMC framework using \texttt{emcee} \citep{emcee}.

We opted to use stochastically-driven damped simple harmonic oscillators (SHOs) for modeling stellar variability, as suggested by \cite{ForemanMackey2017}. This has also been used widely on prior analyses of young transiting planets with \tess\ \citep[e.g.][]{Gilbert2022, Thao2024_1224, Wood2023}. It is common to use a mixture of two SHOs (SHOM or SHOMixture), one at the characteristic period and one at half or double that period. However, we found the second SHO was unconstrained, while a single SHO was sufficient to model the stellar variability over the whole light curve.

We fit for 12 parameters in total. For the planet, we fit for the time of inferior conjunction ($T_0$), orbital period ($P$), planet-to-star radius ratio ($R_P/R_*$), and impact parameter ($b$). We additionally fit for $\sqrt{e}\cos{\omega}$ and $\sqrt{e}\sin{\omega}$ to determine orbital eccentricity ($e$) and the argument of periastron ($\omega$). For the star, we fit for stellar density ($\rho_*$) and two quadratic limb-darkening coefficients ($q_1$ and $q_2$) following the triangular sampling prescription \citep{Kipping2013}. The remaining parameters were used in the GP model: the undamped period of the oscillator ($P$), the damping timescale ($\tau$) and the standard deviation of the process ($\sigma$).

Most parameters evolved under uniform priors with only physical limitations, with the exception of $\rho_*$ which evolved under a Gaussian prior for one of the two fits (see Table \ref{tab:priors}). Transit duration is affected by both the stellar density and orbital eccentricity \citep{Van-Eylen2015}. We initially ran the fit assuming a circular orbit (locking $e=0$) and allowing $\rho_*$ to float (with a lower limit of 0). We found the stellar density disagreed with the expected value from the stellar mass and radius ($1.43\pm0.37$ $\rho_\odot$), suggesting the orbit is eccentric. We opted to rerun the fit letting eccentricity float and placing the Gaussian prior on $\rho_*$.

We ran the MCMC for both fits ($e=0$ and $e$-float) with 50 walkers for 100,000 steps and a 20\% burn-in. The total run time was more than 50 times the autocorrelation time, indicating the number of steps was sufficient for convergence \citep{goodman2010}. 

\begin{figure}
    \centering
    \includegraphics[width=0.98\linewidth]{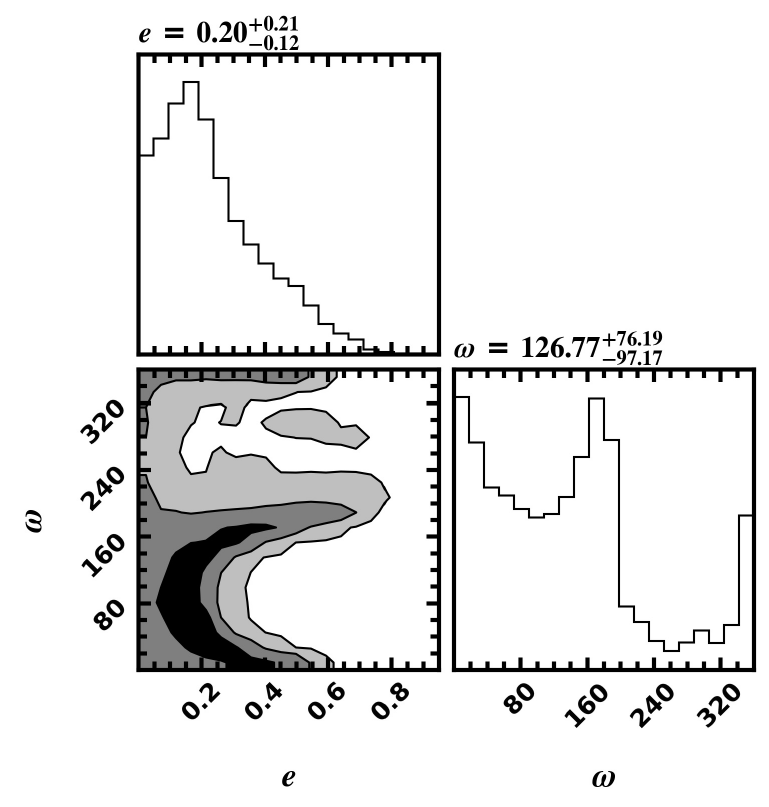}
    \caption{Corner plot of the eccentricity ($e$) and argument of periastron ($\omega$) fit from the \tess\ data. Though the eccentric fit is preferred, the $e$ posterior agrees with a circular orbit.}
    \label{fig:ecc_corner}
\end{figure}

We show the phase-folded light curve and representative section of the light curve and model in Figure \ref{fig:lc} and list the best-fit parameters in Table \ref{tab:parameters}. For both fits, all parameters agreed with one another with the exception of the stellar density. Due to the high disagreement between the stellar density from the transit fit and the stellar density from the mass and radius, we prefer the eccentric transit fit (see Figure \ref{fig:ecc_corner}).

\begin{table*} 
    \caption{\tess\ Transit Parameters of TOI-6448\,b} 
    \begin{tabular}{lccc}
        \hline 
        \hline
        Description & Parameter & \multicolumn{2}{c}{Value}\\ 
        \hline 
        & & e float (preferred) & e = 0\\
        \hline
        \multicolumn{4}{c}{Measured Planet Parameters} \\ 
        \hline
        time of inferior conjunction & $T_0$ (BJD-2457000) & $2214.6125 \pm 0.0026$ & $2214.6127^{+0.0025}_{-0.0024}$\\ 
        orbital period & $P$ (days) & $14.844281^{+3.1\times10^{-5}}_{-3.4\times10^{-5}}$  & $14.844281^{+3\times10^{-5}}_{-3.2\times10^{-5}}$\\ 
        planet-to-star radius ratio & $R_P/R_{\star}$ & $0.0896^{+0.0049}_{-0.005}$ & $0.088^{+0.0044}_{-0.0043}$\\ 
        impact parameter & $b$ & $0.09^{+0.55}_{-0.7}$  & $-0.02^{+0.5}_{-0.48}$\\
        eccentricity parameter & $\sqrt{e}\sin\omega$ & $0.15^{+0.19}_{-0.23}$  & $\cdots$ \\ 
        eccentricity parameter & $\sqrt{e}\cos\omega$ & $-0.01^{+0.44}_{-0.49}$ & $\cdots$ \\
        \hline 
        \multicolumn{4}{c}{Stellar Parameters} \\ 
        \hline
        stellar density & $\rho_{\star}$ ($\rho_{\odot}$) & $1.48^{+0.35}_{-0.36}$ & $3.61^{+0.89}_{-1.0}$\\ 
        limb-darkening coefficient & $q_{1}$ & $0.43^{+0.34}_{-0.27}$ & $0.43^{+0.34}_{-0.25}$\\ 
        limb-darkening coefficient & $q_{2}$ & $0.24^{+0.17}_{-0.16}$ & $0.24 \pm 0.16$ \\ 
        \hline 
        \multicolumn{4}{c}{GP Parameters} \\ 
        \hline
        standard deviation & $\sigma$ & $0.0713^{+0.0048}_{-0.0039}$ & $0.0713^{+0.0048}_{-0.004}$ \\ 
        period & $P$ (days) & $2.662^{+0.078}_{-0.073}$ & $2.666^{+0.077}_{-0.073}$ \\ 
        damping timescale & $\tau$ (days) & $2.9^{+0.35}_{-0.18}$ & $2.89^{+0.33}_{-0.18}$\\ 
        \hline 
        \hline 
        \multicolumn{4}{c}{Derived Parameters} \\ 
        \hline
        semi-major axis to stellar radius ratio & $a/R_{\star}$ & $31.7^{+2.7}_{-3.1}$ & $39.0^{+3.0}_{-5.5}$ \\ 
        inclination & $i$ ($^{\circ}$) & $89.8^{+1.4}_{-1.1}$ & $90.02^{+0.75}_{-0.77}$  \\  
        transit duration (first to fourth contact) & $T_{14}$ (days) & $0.139^{+0.052}_{-0.026}$ & $0.1247^{+0.0058}_{-0.0047}$\\ 
        planet radius & $R_P$ ($R_J$) & $0.782^{+0.074}_{-0.075}$ & $0.768 \pm 0.071$ \\
            & $R_P$ ($R_\oplus$) & $8.77^{+0.83}_{-0.84}$ & $ 8.61 \pm 0.80$\\
        semi-major axis & $a$ (AU) & $0.132^{+0.015}_{-0.017}$ & $0.163^{+0.018}_{-0.026}$ \\ 
        equilibrium temperature$^\dagger$ & $T_{\mathrm{eq}}$ (K) & $742.0^{+38.0}_{-33.0}$ & $669.0^{+49.0}_{-27.0}$\\  
        eccentricity & $e$ & $0.2^{+0.21}_{-0.12}$ & $\cdots$ \\ 
        argument of periastron & $\omega$ ($^{\circ}$) & $127.0^{+76.0}_{-97.0}$ & $\cdots$ \\ 
        \hline 
        \multicolumn{4}{l}{$\dagger$ assuming zero albedo}
    \end{tabular}
    
    \label{tab:parameters}
\end{table*} 

\subsection{Ground-based Transit Fit}

\begin{figure*}
    \includegraphics[width=.36\textwidth]{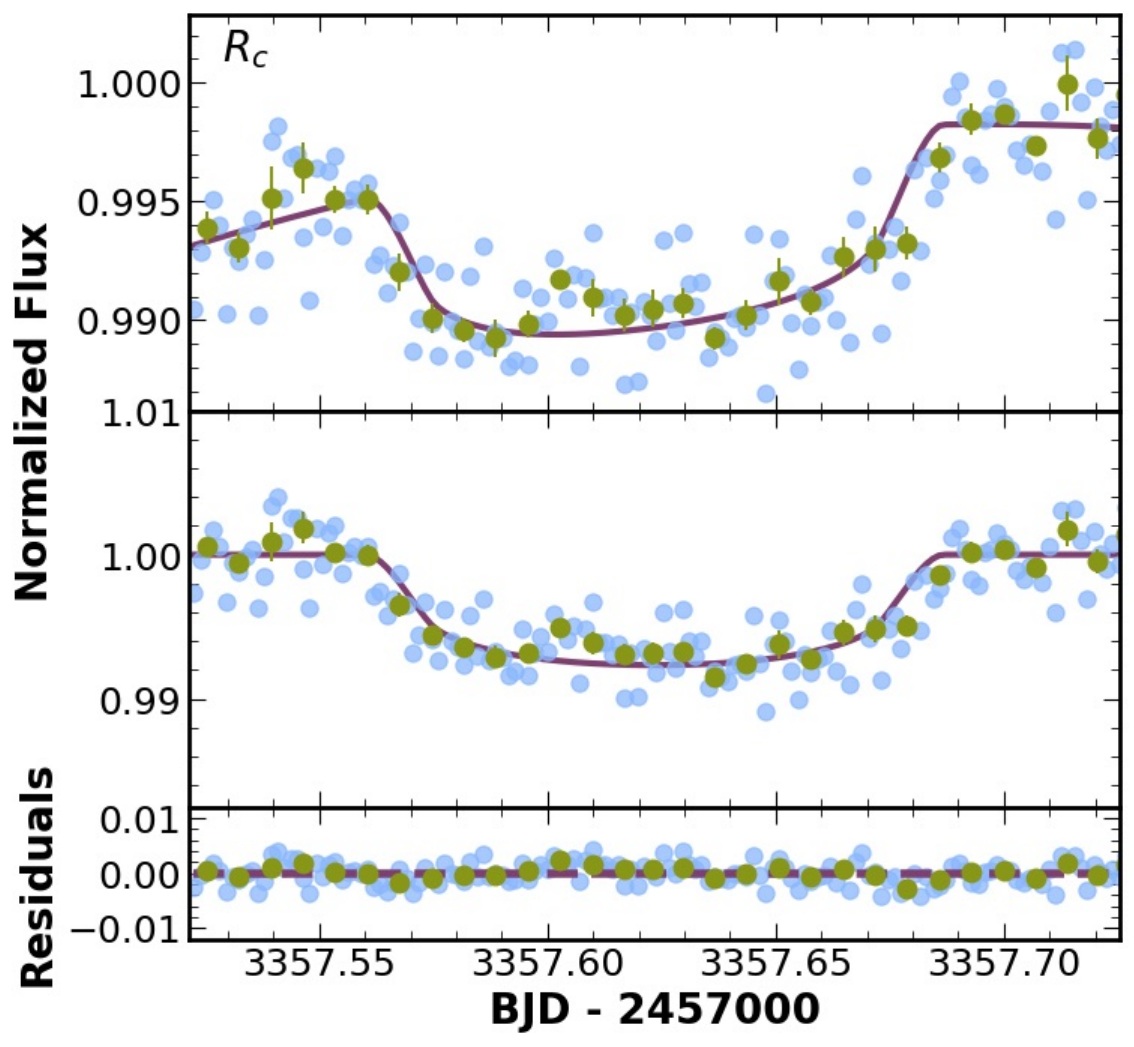}
    \includegraphics[width=.36\textwidth]{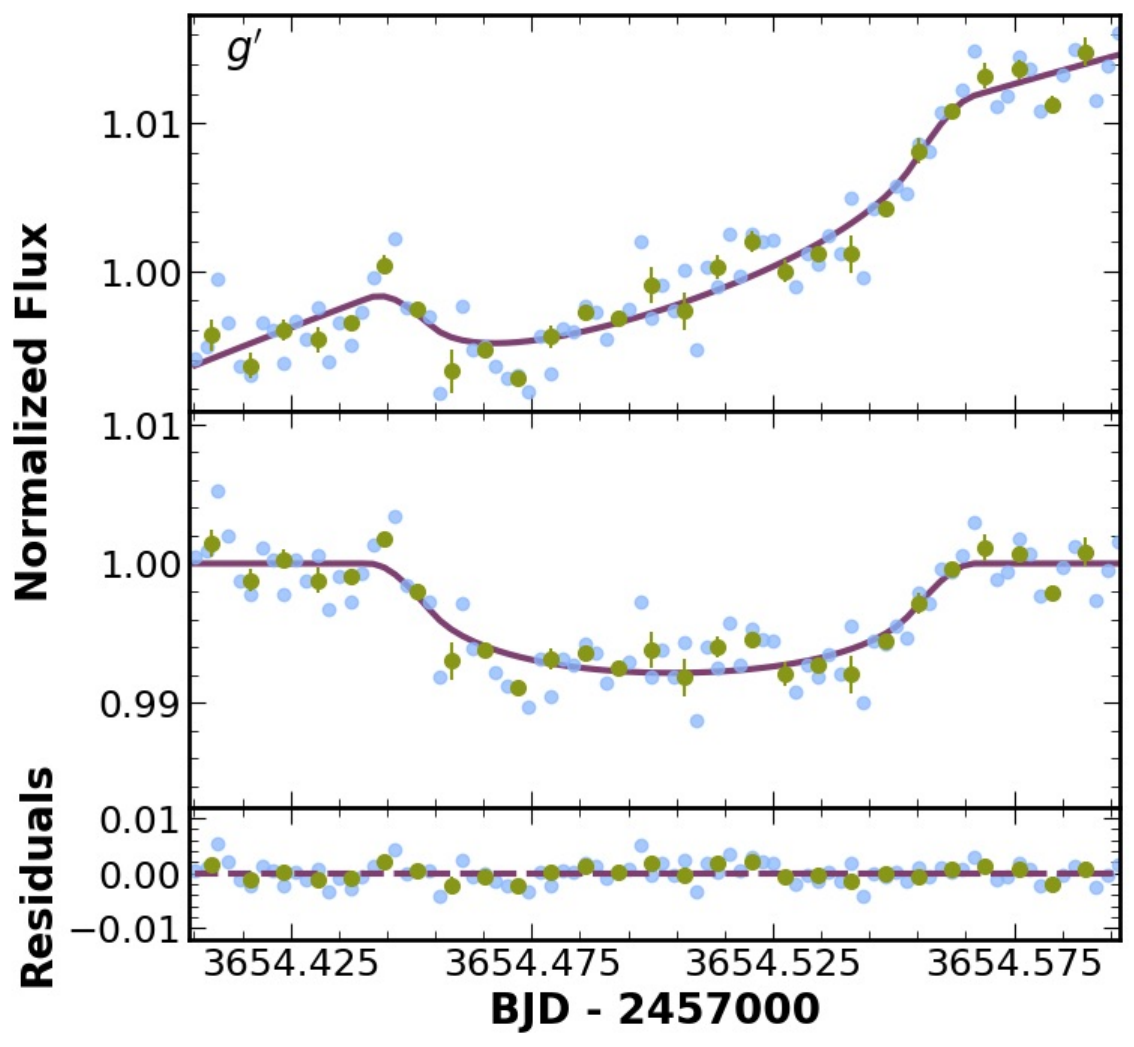}
    \includegraphics[width=.25\textwidth]{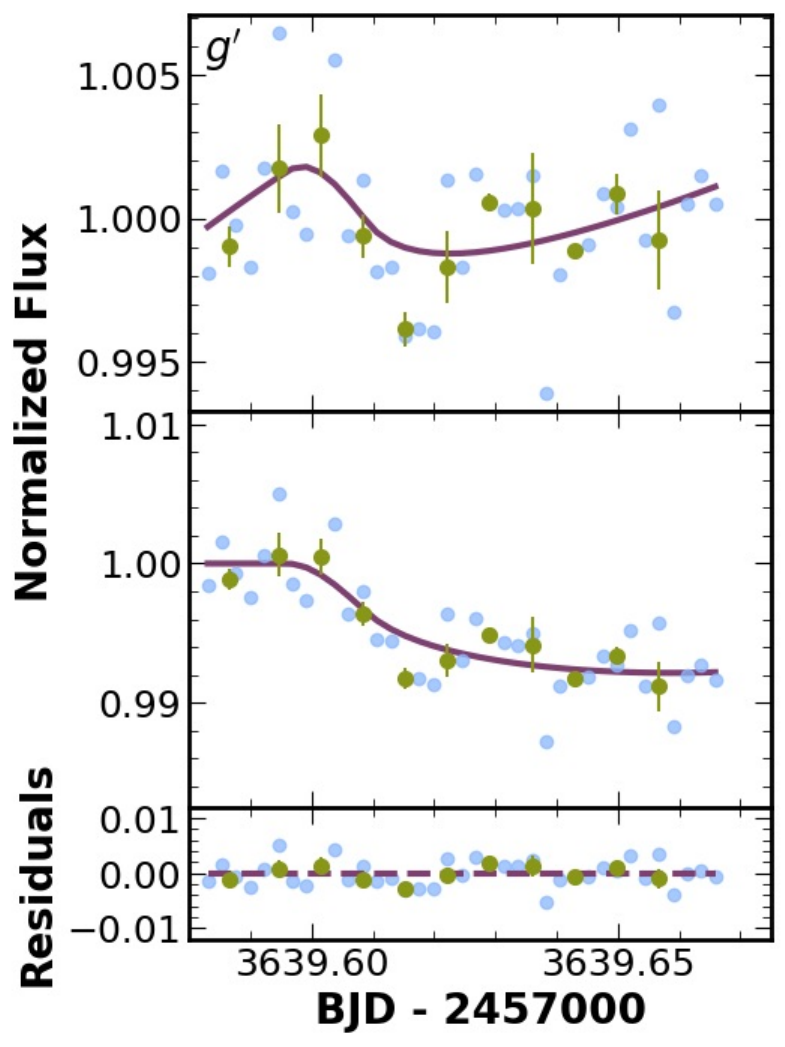} \\
    \includegraphics[width=.48\textwidth]{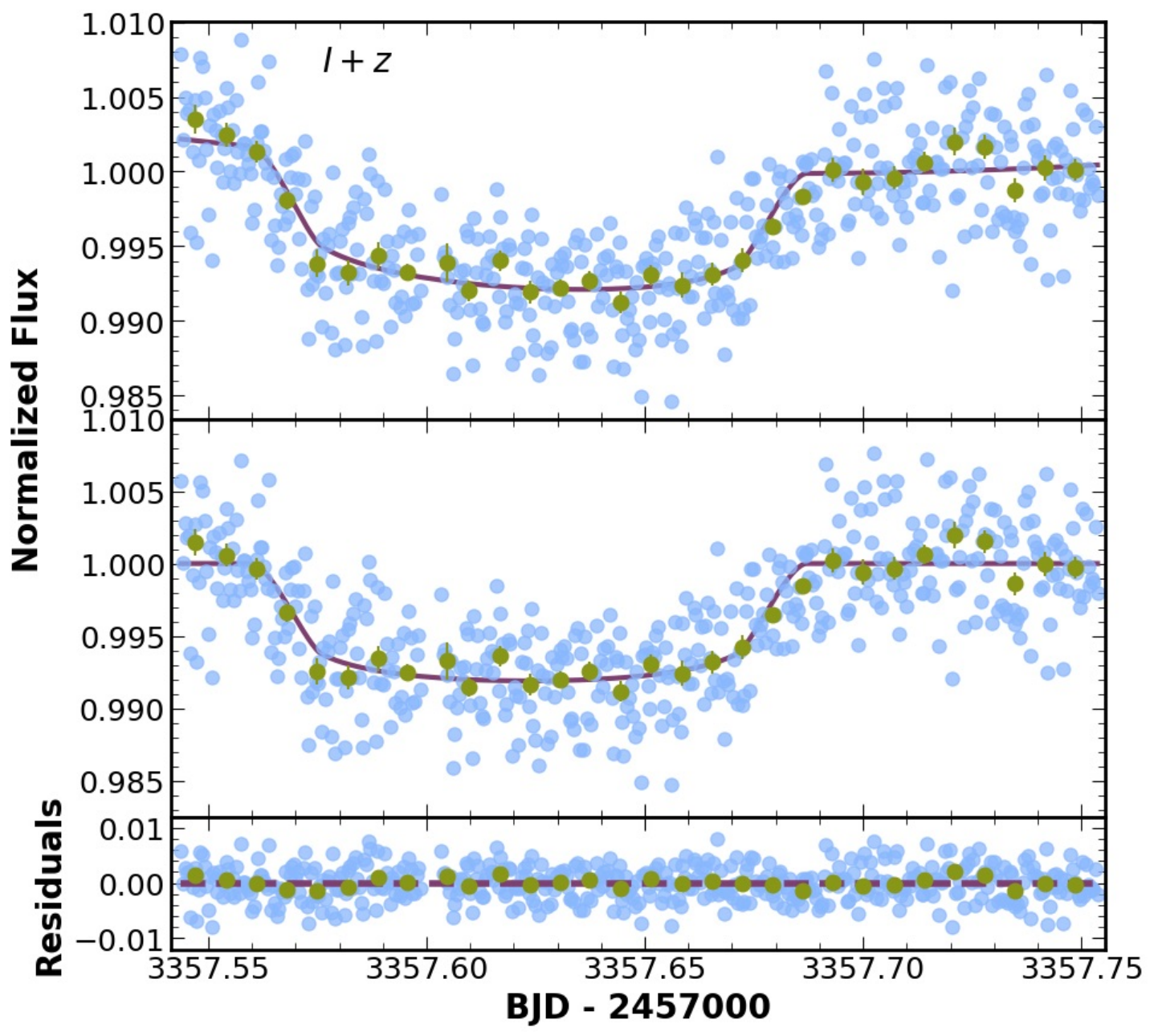}
    \includegraphics[width=.485\textwidth]{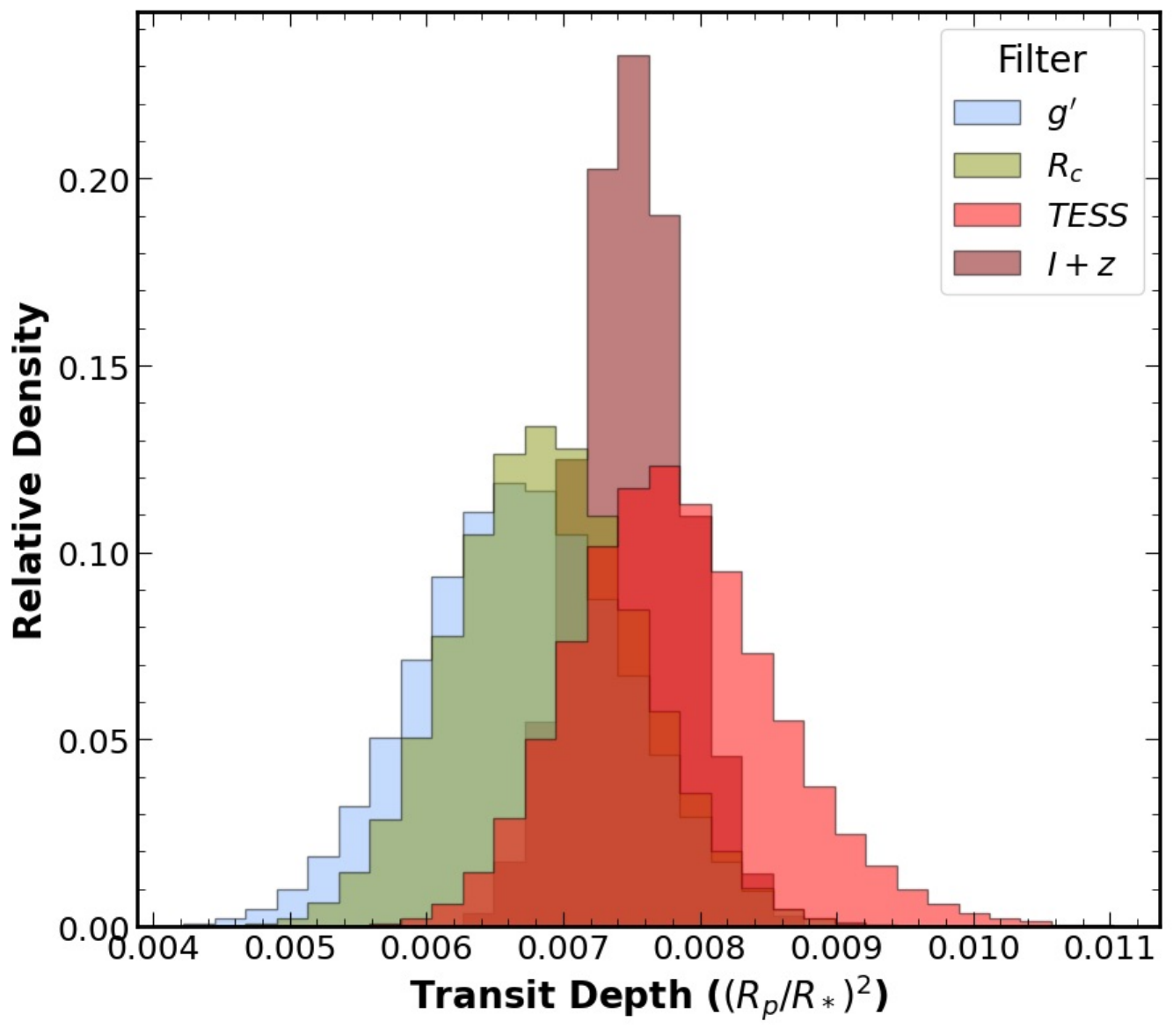}
    \caption{Ground-based light curves (blue) binned to 10 minute intervals (green) for clarity. The transit model is shown in purple. In each set, the top plot is the raw data, the middle plot has the stellar variability model removed, and the bottom plot is the residuals. Bottom right: Transit depth, measured by $(R_p/R_*)^2$, posterior from each transit fit. }
    \label{fig:ground}
\end{figure*}

\begin{table*}
    \centering
    \caption{Ground-based Transit Parameters of TOI-6448\,b} 
    \begin{tabular}{lccccc}
        \hline 
        \hline
        Description & Parameter & \multicolumn{4}{c}{Value}\\ 
        \hline 
        & & $R_c$ & $I+z$ & \multicolumn{2}{c}{$g'$}\\
        \hline
        \multicolumn{6}{c}{Planet Parameters}\\
        \hline
        planet-to-star radius ratio & $R_P/R_{\star}$ & $0.0829^{+0.0040}_{-0.0041}$ & $0.0866\pm{0.0022}$ & \multicolumn{2}{c}{$0.0818^{+0.0047}_{-0.0045}$} \\
        time of inferior conjunction & $T_0$ (BJD-2457000) & \multicolumn{2}{c}{$3357.62350^{+0.00079}_{-0.00084}$} & $3639.6568^{+0.0029}_{-0.0028}$ & $3654.5041\pm0.0012$  \\
        \hline
        \multicolumn{6}{c}{Variability Parameters}\\
        \hline
        constant & a & $-0.00685\pm0.00085$ & $0.00218\pm0.00040$ & $-0.0030\pm0.0010$ & $-0.00639\pm0.00065$\\
        linear coefficient & b & $0.059\pm0.025$ & $-0.033\pm0.011$ & $0.156\pm0.065$ & $0.131\pm0.026$\\
        quadratic coefficient & c & $-0.17\pm0.12$ & $0.117\pm0.055$ & $-0.54\pm0.72$ & $-0.11\pm0.14$\\
        \hline

    \label{tab:ground}
    \end{tabular}
\end{table*}

We fit the ground-based photometry separate from the \tess\ data. Our main goal with the ground-based data is to confirm the transit timing and depth are consistent and that the transit is occurring on-source (i.e., for false-positive assessment). 

Two of the ground-based transits ($R_C$ and $I+z$ filters) were taken simultaneously. This is ideal for checking the transit depth as we do not have to worry about changing spot properties or other effects that vary on week-to-month timescales, and let us lock to a common $T_0$. 

We fit the photometry using a \texttt{BATMAN} transit model and a second order polynomial to handle stellar variability. The baseline for the photometry was too narrow to train a periodic GP, and the photometry (including the \tess{} photometry) is well described by a simple polynomial over windows of $<0.5$ days (Figure~\ref{fig:lc}). The second-order polynomial had the form
\begin{equation}
    f_{corrected} = f_{raw} - (a + b\times t + c\times t^2),
\end{equation}
where $t$ is the number of days since the first datapoint (to keep all values positive).

Since we are only interested in constraining the transit depth and timing, we lock all planet parameters to the \tess\ fit with the exception of $R_P/R_*$ and $T_0$. We locked limb-darkening coefficients for each filter using values determined with \texttt{LDTK} \citep{Parviainen2015}. For the simultaneous transits, we fit the transits with the same $T_0$ but independent $R_P/R_*$ and variability models (to account for differences in telescope systematics and data reduction). The two $g'$ transits were fit independently of the two simultaneous transits and fit using a common $R_P/R_*$ but separate $T_0$ and variability parameters.

As above, we ran the MCMC using 50 walkers with 50,000 steps and a 20\% burn-in. We show the resulting fit in Figure \ref{fig:ground} and present the best-fit parameters in Table \ref{tab:ground}. As can be seen in Figure \ref{fig:ground}e, all transit depths are consistent with each other at 1$\sigma$.

\subsection{Global analysis}

To achieve the most precise transit parameters, we re-ran our \texttt{MISTTBORN} transit fit using the \tess\ and ground-based photometry. We opted to use the same priors and GP kernel as the \tess-only fit (Section \ref{sec:tessFit}), with the exception of an additional Gaussian prior to all limb-darkening coefficients ($\pm 0.1$).

We model variability in both \tess{} and ground-based data with the same SHO kernel. This works if the dominant source of variability in each dataset is the rotation signal. While the ground-based data also contains instrumental and atmospheric effects, in practice these are small compared to the variability signal (Figure \ref{fig:lc}), and the data gaps allow the GP to adjust to the newer dataset. One GP does, however, prevent us from using simultaneous datasets, so we removed the $R_c$ data in favor of the more precise $I+z$ data.

We fit for 16 parameters in total. For the four datasets (\tess, $I+z$, and two $g'$), we use six common parameters to describe the transit ($T_0$, period, $R_P/R_*$, $b$, $\sqrt{e}\sin{\omega}$, and $\sqrt{e}\cos{\omega}$), three common parameters to describe the SHO GP kernel (period, $\tau$, and $\sigma$), and a common stellar density ($\rho_*$). For each filter, we fit for two limb darkening coefficients.

We ran the MCMC with 50 walkers for 100,000 steps and a 20\% burn-in. We present the best-fit parameters in Table \ref{tab:global}. While the ground-based data is relatively precise, there is far more \tess{} data, and the ground-based data did not significantly improve the baseline (which would normally provide significant improvements in $P$). Thus, we find only a slight improvement to the $T_0$, period, and $R_P/R_*$ parameters compared to the \tess-only fit. 

\begin{table*}[]
    \centering
    \caption{Global Transit Parameters of \planetname}
    \begin{tabular}{lcc}
    \hline
    \hline
    Description & Parameter & Value\\ 
        \hline 
        \multicolumn{3}{c}{Measured Planet Parameters} \\ 
        \hline
        time of inferior conjunction & $T_0$ (BJD-2457000) & $2214.6136^{+0.0024}_{-0.0023}$ \\
        orbital period & $P$ (days) & $14.844261^{+2.7\times10^{-5}}_{-2.8\times10^{-5}}$ \\ 
        planet-to-star radius ratio & $R_P/R_{\star}$ & $0.0866\pm0.0038$ \\ 
        impact parameter & $b$ & $0.51^{+0.14}_{-0.95}$ \\
        eccentricity parameter & $\sqrt{e}\sin\omega$ & $0.14^{+0.17}_{-0.23}$ \\ 
        eccentricity parameter & $\sqrt{e}\cos\omega$ & $-0.17^{+0.54}_{-0.47}$ \\
        \hline 
        \multicolumn{3}{c}{Stellar Parameters} \\ 
        \hline
        stellar density & $\rho_{\star}$ ($\rho_{\odot}$) & $1.48^{+0.36}_{-0.37}$ \\ 
        \tess\ limb-darkening coefficient & $q_{1,\tess}$ & $0.452^{+0.094}_{-0.092}$ \\ 
        \tess\ limb-darkening coefficient & $q_{2,\tess}$ & $0.219^{+0.098}_{-0.090}$ \\ 
        $g'$ limb-darkening coefficient & $q_{1,g'}$ & $0.931^{+0.047}_{-0.071}$ \\ 
        $g'$ limb-darkening coefficient & $q_{2,g'}$ & $0.265^{+0.093}_{-0.091}$ \\ 
        $I+z$ limb-darkening coefficient & $q_{1,I+z}$ & $0.597\pm0.094$ \\ 
        $I+z$ limb-darkening coefficient & $q_{2,I+z}$ & $0.255^{+0.091}_{-0.097}$ \\ 
        \hline 
        \multicolumn{3}{c}{GP Parameters} \\ 
        \hline
        standard deviation & $\sigma$ & $0.0699^{+0.0045}_{-0.0037}$ \\ 
        period & $P$ (days) & $2.649^{+0.078}_{-0.070}$ \\ 
        damping timescale & $\tau$ (days) & $2.83^{+0.29}_{-0.16}$ \\ 
        \hline 
        \hline 
        \multicolumn{3}{c}{Derived Parameters} \\ 
        \hline
        semi-major axis to stellar radius ratio & $a/R_{\star}$ & $31.5^{+2.4}_{-2.1}$ \\ 
        inclination & $i$ ($^{\circ}$) &  $88.95^{+1.96}_{-0.34}$ \\ 
        transit duration (first to fourth contact) & $T_{14}$ (days) & $0.147^{+0.073}_{-0.028}$ \\ 
        planet radius & $R_P$ ($R_J$) & $0.756\pm0.068$ \\
            & $R_P$ ($R_\oplus$) & $8.47\pm0.76$\\
        semi-major axis & $a$ (AU) & $0.131\pm 0.014$ \\ 
        equilibrium temperature$^\dagger$ & $T_{\mathrm{eq}}$ (K) & $745.0^{+28.0}_{-31.0}$ \\ 
        eccentricity & $e$ & $0.2^{+0.26}_{-0.12}$ \\ 
        argument of periastron & $\omega$ ($^{\circ}$) & $143.9^{+52.1}_{-100.0}$ \\ 
        \hline 
        \multicolumn{3}{l}{$\dagger$ assuming zero albedo}
    \end{tabular}
    
    \label{tab:global}
\end{table*}

\section{Injection-Recovery Analysis}\label{sec:injrec}

We performed an injection-recovery analysis to determine our sensitivity to additional planets in the system and the reliability of the detection of TOI-6448\,b. 

\begin{figure}
    \centering
    \includegraphics[width=0.98\linewidth]{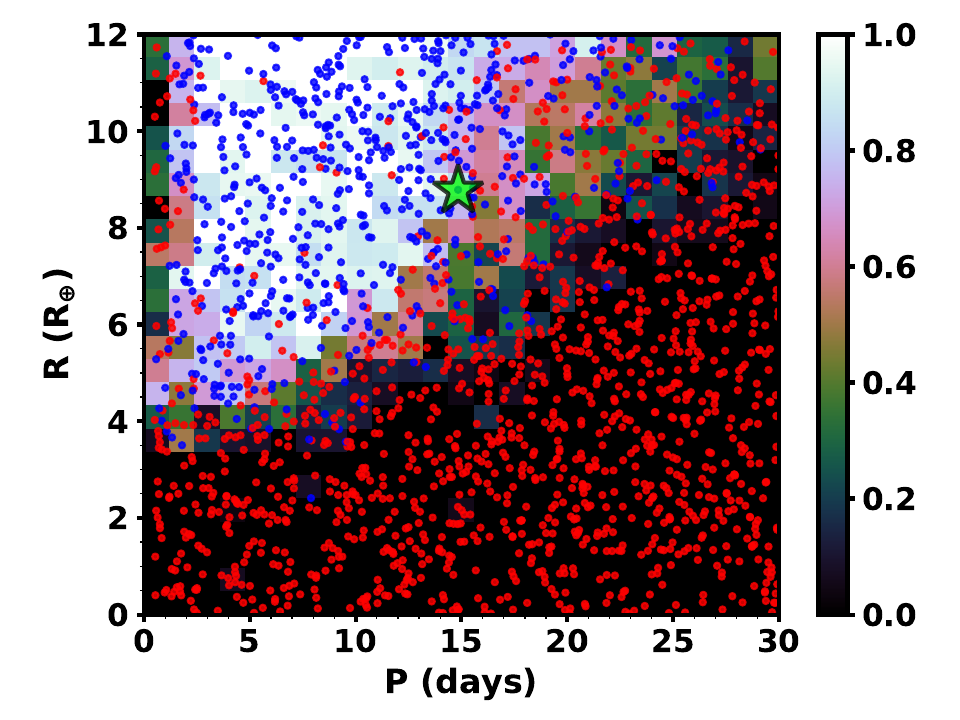}
    \caption{Injection-recovery map in period-radius space for TOI-6448. The green star marks TOI-6448\,b. Blue points indicate recovered planets and red points are planets which were not recovered. Only 20\% of the individual injected signals are shown for clarity. The background is color-coded by overall completeness for a given bin. }
    \label{fig:injrec}
\end{figure}

We injected 10,000 randomly generated synthetic planet signals into the raw light curve extracted in Section \ref{sec:tessobs}. Planets were given a radius 0.01--12 $R_\oplus$, period 0.5--30 days, and $T_0$ between the start of the first sector and one period away. For simplicity, we restrict eccentricity to 0. We then re-ran \texttt{Notch} to attempt to recover the signal. A signal is considered ``recovered" if \texttt{Notch} identifies the correct planet period and $T_0$ (allowing for a 1\% deviation from the input on either) at $\ge8\sigma$ significance. We show the resulting distribution in Figure \ref{fig:injrec}. TOI-6448\,b lands in a region with $\sim80\%$ recoverability. 

\section{False Positive Analysis} 
\label{sec:fp}

Using \texttt{TRICERATOPS} \citep{triceratops}, we initially calculated a false positive probability FPP $\sim 12\%$. \texttt{TRICERATOPS} favored TIC 320411046 as the source of a nearby eclipsing binary (NEB) scenario. The ground-based transits exclude this source (separation\,$\simeq$\,25\arcsec), ruling out this scenario (see Section~\ref{subsec:lcogt} and text below). Recalculating the FPP with this NEB excluded, we find a FPP $<10^{-5}$, statistically validating the signal as a transiting planet orbiting \starname.

A typical statistical false-positive assessment (such as with \texttt{TRICERATOPS}) for young planets can be complicated by young stellar morphology and the puffy planet size. Large ($5-12R_\oplus$) planets are rare around old stars, but common around $<$200\,Myr systems \citep[e.g.][]{Fernandes2022, Vach2024_ocr}, so the priors will significantly overestimate the false-positive probability. Further, \texttt{TRICERATOPS} and similar codes require detrended data, and the results on highly variable stars are sensitive to the quality of the detrending. So we also statistically validate the planet using the combination of follow-up data and the overall properties of the light curve. We handle the potential sources of false positives one at a time below.

{\bf Stellar variations:} the transit depth is consistent over 4-years and across the four wavelengths to $<1\sigma$ (\tess, $g'$, $R_c$, and $I+z$). This excludes stellar signals like spots or flares that change over these timelines and with wavelength. 

{\bf An eclipsing system orbiting \starname:} an eclipsing system should show V-shaped transits, particularly at redder bands where the contrast ratio is more favorable. More importantly, at 35\,Myr, a brown dwarf or low-mass star will be pre-main-sequence and have a radius of $>15R_\oplus$ \citep{Marley2021}. Such a star would emit negligible flux compared to the primary in our bluest filter (so the inferred depth is reflective of the size of the eclipsing object). Our transit fit is $7\sigma$ below this. 

{\bf An unseen star (bound or background):} the last possible scenario is that the transit signal is associated with another star, either a bound or unassociated eclipsing or transiting system. We can provide a range of constraints on any unseen object from the transit, spectra, and archival imaging. 

The transit shape and depth sets the magnitude limit for the faintest companion that could cause the transit. Following \cite{Vanderburg2019}, we find the companion would need to be $\Delta T < 1.92$ mags at 95\% confidence. 

We can also set color limits on any possible companion using the transit depths from the four wavelengths \citep{THYMEV_FF}. A stellar companion would cause the transit depth to vary across wavelength as the companion's relative contribution changes with wavelength. Using the 95th percentile of the distribution of the transit depth ratio, we find that the $T-I+z$ color sets the tightest limit, requiring the companion to be $<0.2$\,mags redder than the primary.

The transit is recovered in ground-based imaging with apertures as small as 3.7\arcsec{}. This easily excludes the main source of false positives from \texttt{TRICERATOPS}.
No source is detected in \gaia{} imaging within that region, which can detect stars down to $G\simeq20$ within 0.8\arcsec{} of the primary \citep{Ziegler2018}. \gaia{} is sensitive to any star that satisfies the color and magnitude constraints above; the only remaining option is for a source that is unresolved in \gaia. 

\begin{figure}
    \centering
    \includegraphics[width=.48\textwidth]{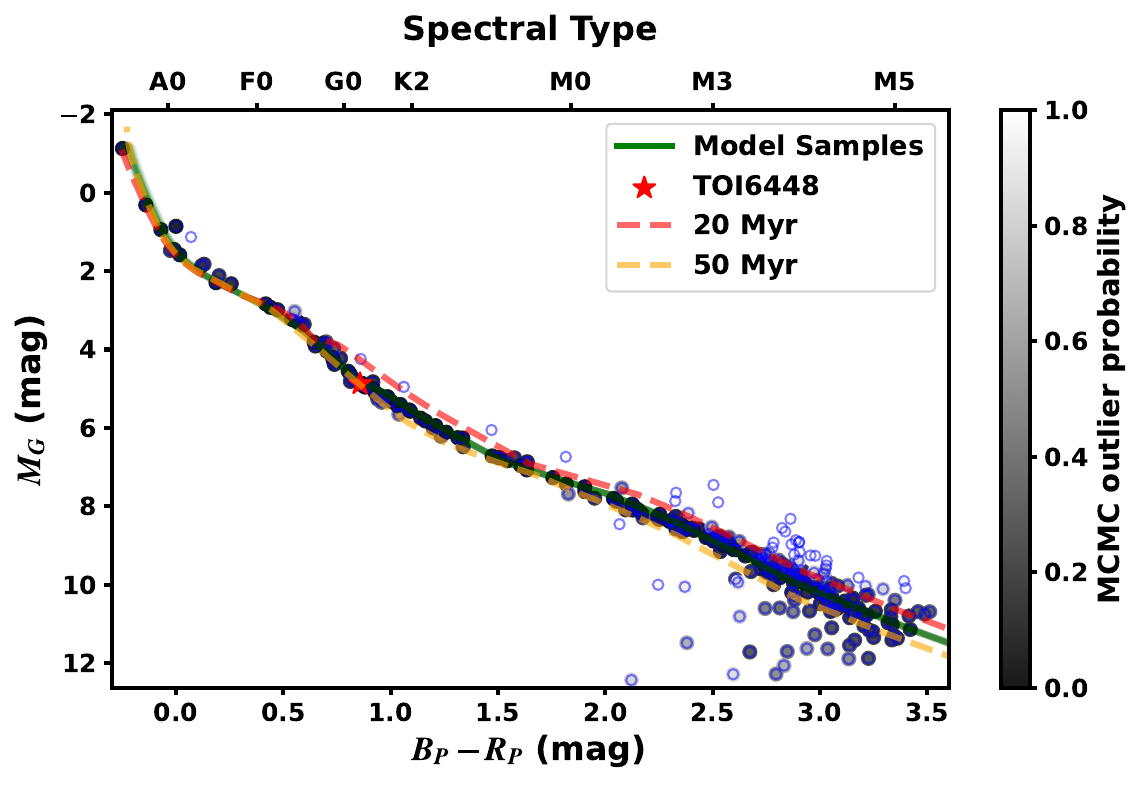}
    \caption{Color-magnitude diagram of stars $<$35\,pc and tangential velocity $<$1\kms\ away from \starname{} (blue circles) compared to model predictions from PARSEC. Each target is shaded based on the probability it is part of the single-star single-age population. Green lines show 100 random samples from the MCMC, and an older (orange dashed) and younger (red dashed) model are shown for comparison. The age is tightly constrained by the handful of BA stars, the pre-main-sequence M dwarfs, and the lack of pre-main-sequence G dwarfs. Many of the stars above the sequence are consistent with being binaries, while many below the sequence are likely non-members or stars with low SNR BP photometry.}
    \label{fig:cmd}
\end{figure}

We can set limits on the brightness of any unresolved star using the population's CMD (see Figure \ref{fig:cmd}). A well-populated single-star and single-age CMD stars of a given color occupy a narrow locus with a measured vertical scatter $\sigma_M$ in absolute magnitude. Any unresolved companion adds flux to the combined system, and the contrast between the primary and companion ($\Delta m$) satisfy:
\begin{equation}  
  \Delta m
  \;\ge\;
  -2.5\,\log_{10}\!\bigl[\,10^{\Delta_{M,lim}/2.5}-1\bigr],
\end{equation}
where $\Delta_{M,lim}$ is the maximum difference between the predicted absolute magnitude and the observed one, accounting for measurement uncertainties and intrinsic scatter in the CMD. Effectively, any unseen companion with a lower contrast than $\Delta m$ would be detected as a higher-than-expected CMD position.

The offset $\Delta_{M,lim}$ can be estimated using a model isochrone, but that is subject to systematics in the model. Instead, we use similar stars (within 1 mag in $M_G$) within the same population, cutting out those with RUWE$>1.2$ (likely binaries). This will be conservative, as the sample will contain some some real binaries. We then perturb individual star $M_G$ and $B_P-R_P$ colors according to their photometric and parallax uncertainties, and interpolate their perturbed colors to predict \starname's $M_G$ given it's (perturbed) color. We take the 99th percentile limit as $\Delta_{M,lim}$, which yields a corresponding contrast of $\Delta G>2.4$ for the unseen companion. This limit is high in part because \starname{} sits slightly below the sequence of nearby stars (Figure~\ref{fig:cmd}). No bound star is consistent with both the color and brightness limits from the CMD, transit shape, and chromaticity. 

The final test is the lack of a second set of lines in our follow-up spectra. Using the SNR of the reddest order of our MIKE spectra, we expect to detect a second set of lines as faint as $\Delta m$ of 5.82 within 1\arcsec of the source (the slit size). Although this assumes similar-strength spectral features as the primary, low rotational broadening and a velocity offset from the target. Under this false-positive scenario, the background star must also be an eclipsing system, so it should undergo velocity variations that would be likely to appear in the CHIRON spectrum. However, since the parameters of the hypothetical unbound star are unknown, this constraint is only suggestive, and is not included in the \texttt{MOLUSC} analysis below or the \texttt{TRICERATOPS} analysis above.

To help quantify this effect we use \texttt{MOLUSC} \citep[Multi-Observational Limits on Unseen Stellar Companions;][]{Wood2021_molusc}, which generates synthetic binaries and compares them to observational data. We generated 100,000 synthetic companions to compare to observed radial velocities and \gaia\ imaging. We simulate the finite aperture of the ground-based data by adding a contrast curve that is infinite past 3\arcsec. Of the generated synthetic companions, \texttt{MOLUSC} ruled out 78.05\%. Applying the above \tess\ magnitude cut based on the transit shape and color cut based on the chromaticity of the transit removes an additional 21.62\%. The remaining 335 (0.33\%) binaries are all unresolved SB2 with high mass ratios ($>0.7$) on long period orbits ($>88$yr) that might have velocities similar to the primary. Because the survivors are relatively bright, all 335 are eliminated by the CMD constraints. 

The input velocities are from two different instruments, raising the possibility of offsets, but instrumental offsets are generally $<$100\mps \citep{Katz2019}, which is small compared to the precision here. Further, brown dwarfs or other close-in systems that would be ruled out by RVs are disfavored to the lack of chromaticity, the transit shape, and the predicted transit depth from a 34\,Myr brown dwarf. Re-running our analysis absent the input RVs does not change any conclusions. 

\section{\starname's parent population}

\subsection{Selection of stars co-moving with \starname} 
\label{sec:friendfinder}

We searched for co-moving stars with \starname\ using \texttt{FriendFinder}\footnote{\url{https://github.com/adamkraus/Comove}} \citep{THYMEV_FF}. Using \starname's radial velocity, \texttt{FriendFinder} uses \gaia\ Data Release 3 astrometry to compute the XYZ position and UVW velocities for nearby stars and looks for stars within set velocity and position bounds.

Since our main goal of identifying comoving stars is to calculate a precise age of the cluster, we want to identify stars with high probabilities of being true members with a focus on creating a ``clean'' list rather than a complete one. We also want to ensure we are removing stars that are likely to be parts of other regions of Vela, which may have similar kinematics but not necessarily the same age. To this end, we searched for stars with a tangential velocity within 1 \kms\ and a 3D distance within 40 pc of \starname. This resulted in 393 candidate comoving stars. We list the candidate members in Table \ref{tab:bigTable}.

\begin{figure*}
    \centering
    \includegraphics[width=0.98\linewidth]{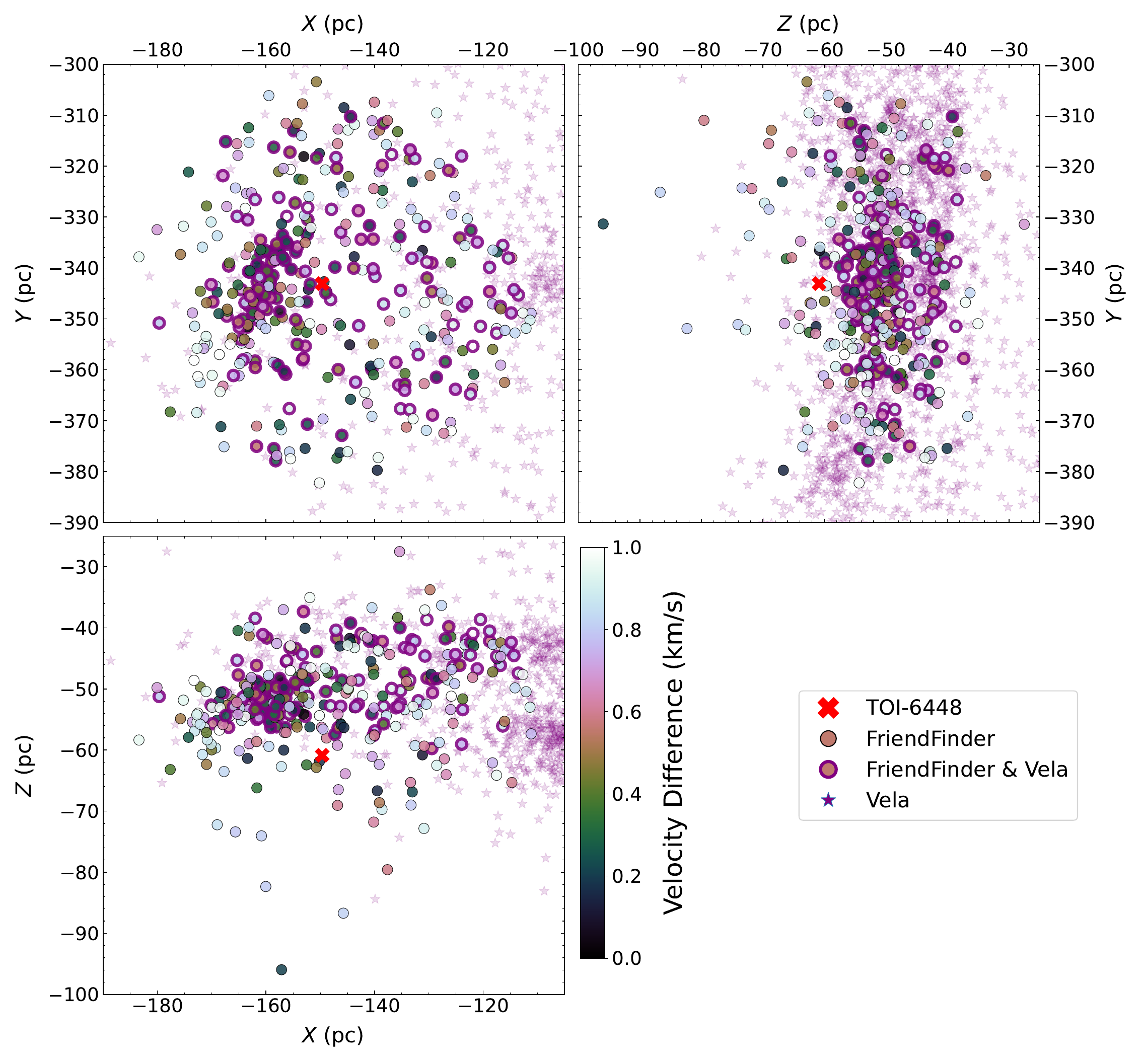}
    \caption{Candidate co-moving stars to TOI-6448 (red x). All stars identified with \texttt{FriendFinder} are shown as the colored circles (colored by their velocity offset to TOI-6448). Stars that are in the same region and identified in \cite{CantatGaudin2019} as part of Vela Population IV are shown in the background as the translucent purple stars. Stars that are in both lists are shown as the colored circles outlined in purple.}
    \label{fig:friendfinder}
\end{figure*}

\subsection{Connection to Vela}

\cite{CantatGaudin2019} clusters \starname\ into Vela Population IV with a 77\% membership probability. Of the 393 stars identified by \texttt{FriendFinder}, 221 overlap with the \cite{CantatGaudin2019} list (see Figure \ref{fig:friendfinder}). Of these, 84\% (185) have a $\geq$90\% membership probability.

Multiple other studies place \starname\ in Collinder 140 \citep[e.g.,][]{Pang2022,vanGroeningen2023,qin2023} with membership probabilities of 90-100\%. However, \cite{CantatGaudin2019} argues that Collinder 140 is part of Vela Population IV, so these studies are all consistent with membership. We will refer to the parent cluster as Vela for the rest of this work. 

As we show below, \starname{} has rotation consistent with the parent population which itself is consistent with an $<$80\,Myr population. From our CHIRON and MIKE spectra, we estimate a lithium equivalent width of 170$\pm$15\,m\r{A} for \starname. This places it solidly within the 20-50\,Myr bin from \citet{Jeffries2023} and solidly rules out ages $>$100\,Myr. As we show below, this is consistent with the age of the parent population (35\,Myr). Combined with the already high membership probabilities in the literature, we conclude that \starname{} is a member Population IV of Vela.

\subsection{Age Analysis}\label{sec:age}

\cite{CantatGaudin2019} suggests Vela Population IV is comprised of five previously known clusters; NGC 2547, NGC 2451B, Collinder 135, UBC 7, and Collinder 140. The reported ages of these clusters range from 27--170\,Myr \citep{Dias2002, Kharchenko2013,Bossini2019}.

Motivated by the age spread and importance of this target for statistical work, we decided to re-analyze the age and confirm \starname's membership into Vela Population IV. To derive the most precise age for \starname\ and Vela IV, we combine the results of isochronal modeling, variability-based aging, and gyrochronology. 

\subsubsection{Isochronal Modeling}\label{sec:isochrone}

We fit the \gaia\ color-magnitude diagram using a mixture model as described in \cite{THYMEVI_1227}. To briefly summarize, we compare the \gaia\ colors and absolute magnitudes ($M_G$) to the combination of a single-star single-age sequence derived from a model isochrone, and a second outlier population. The outlier population may be made of binaries, non-members, targets with poor photometry or parallaxes, or members of other (younger or older) regions of Vela. In principle, one could model this using 3-5 populations in the mixture model instead of just two, e.g., one model targeting member binaries, one capturing field interlopers, another for young Vela interlopers, and a final one for targets with bad measurements. In practice, there are too few stars in these groups to constrain the population parameters.  

The main population is modeled using two parameters - the age ($\tau$) and the overall extinction ($E(B-V)$), while the outlier population is modeled as an offset ($Y_B$) and variance ($V_B$) from the primary population, both measured in magnitude. Additional parameters $f$ and $P_B$ describe the missing scatter (e.g., differential extinction between stars or underestimated measurement uncertainties) and the amplitude of the outlier population (the fraction of stars not in the single-star single-age group). 

We tested models from PARSECv2.0 \citep{Nguyen2022_parsec20} and the Dartmouth stellar evolution program \citep[DSEP;][]{Dotter2008} with magnetic-enhancement \citep[DSEP-mag;][]{Feiden2016}. The results were consistent at $\simeq2\sigma$, with PARSEC yielding a lower age (31.5$\pm$\,1.5Myr) than the DSEP-mag models (38$\pm$3\,Myr). Both fit the overall CMD well, although the DSEP-mag models do not cover the warmest (spectral class B and A) stars in the sample, yielding larger uncertainties

As can be seen in Figure~\ref{fig:cmd}, the age of the population around \starname{} is tightly constrained by the handful of BA stars (which evolve quickly), the population of pre-main-sequence M dwarfs, and the lack of pre-main-sequence G and early-K dwarfs (which rules out significantly younger ages). The combination also means the results are weakly sensitive to assumptions about metallicity; adjustments of $\pm0.2$~dex change the age by $\simeq$2\,Myr. Including this and accounting for both grid results we adopted a more conservative age of 34$\pm$3\,Myr with a mean reddening of $E(B-V)=0.05\pm0.01$.

\subsubsection{Variability age}
Taking advantage of variable stars having higher photometric uncertainties and the relationship between stellar activity and age, \cite{BarberMann2023} fit a Skumanich-like relationship between age and excess uncertainties in \gaia\ photometry \citep{Riello2021}. We ran \texttt{EVA}\footnote{\url{https://github.com/madysonb/EVA}} (Excess Variability-based Age) to query \gaia, make appropriate quality cuts, and compute the age. Using \texttt{EVA}, we calculate an age of $39^{+10}_{-8}$ Myr taking into account all three bands ($G$, $B_P$, and $R_P$), though the ages from the three bands agreed within 1$\sigma$ to each other.

\subsubsection{Gyrochronology}\label{sec:gyro}

We calculated the rotation periods following \cite{Barber2025_2076e} and \cite{Boyle2025}. To summarize, we downloaded light curves for 298 of the 393 stars in our sample using the \texttt{unpopular} package \citep{2022AJ....163..284H}. Then for each star, for each sector, we used a Lomb-Scargle periodogram with a linearly spaced search grid with 100,000 steps spanning 0.2 to 20 days. If a star was observed in multiple sectors, the rotation period corresponding to the highest power was adopted. We assigned uncertainties on the rotation periods using the empirical relation in \cite{Boyle2025}. 

\begin{figure}
    \centering
    \includegraphics[width=0.98\linewidth]{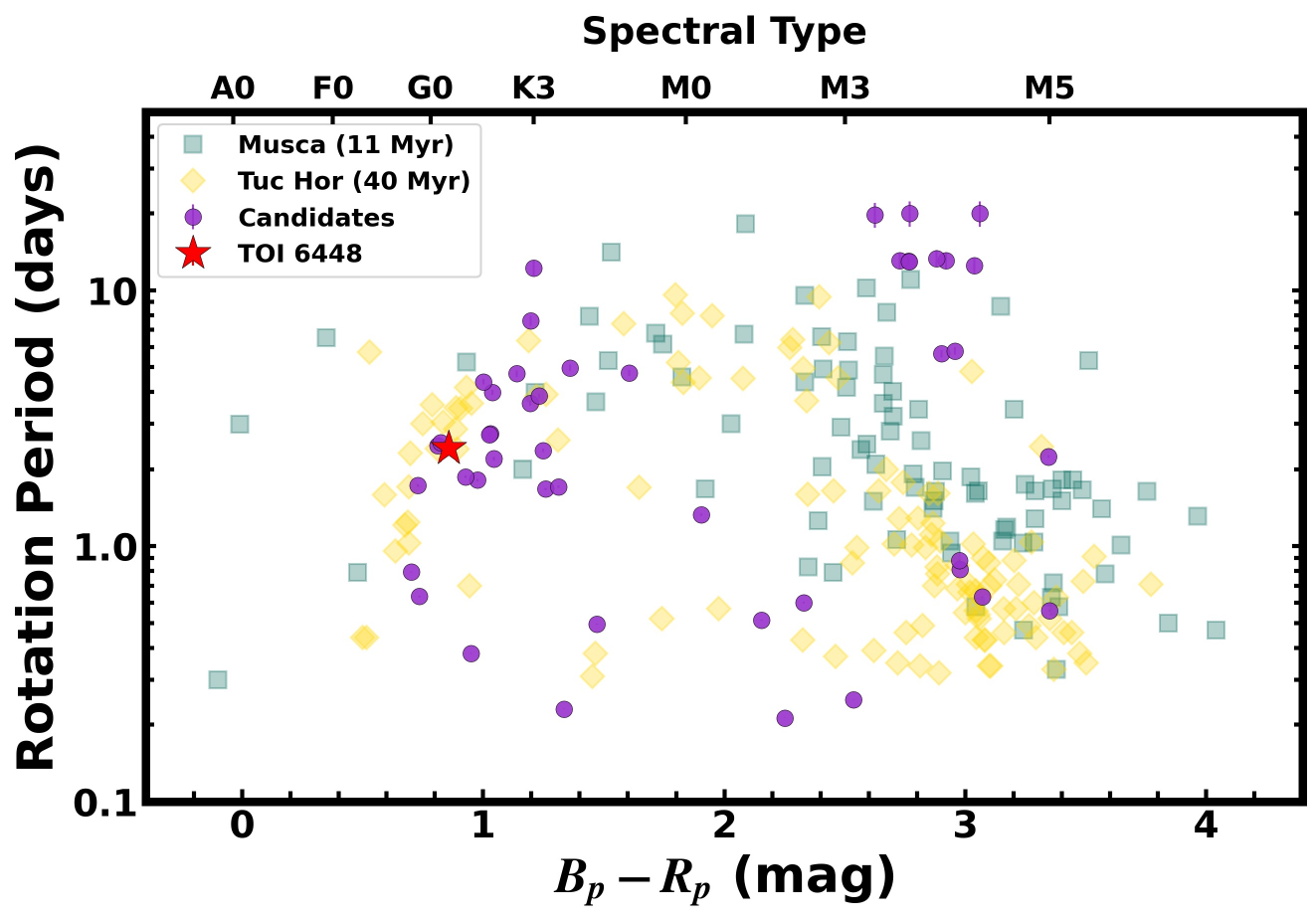}
    \caption{Rotation periods of candidate members (purple circles) comoving with TOI 6448 (red star). Only rotation periods measured to a Lomb-Scargle power $>0.03$ are shown. The rotation periods for Musca \citep[$\sim$11 Myr;][]{THYMEVI_1227} and Tuc-Hor \citep[$\sim$40 Myr;][]{Gagne2020} (green squares and yellow diamonds, respectively) are shown for reference.}
    \label{fig:prot}
\end{figure}

We find TOI-6448 to have a rotation period of $2.412\pm0.037$ days. Compared to the remaining candidate members and members of similarly aged young populations (Figure \ref{fig:prot}), the rotation period of \starname\ is consistent with membership.

We attempted to convert the rotation periods to an age estimate using \texttt{gyro-interp} \citep{Bouma2023}. Due to stars spinning-up during their first $\sim100$\,Myr, \texttt{gyro-interp} is calibrated only for stars $>$80\,Myr. We use the rotation periods to confirm group membership, but gyrochronology only sets an upper limit on the age.

\subsubsection{Combining Age Estimates}
\begin{figure}
    \centering
    \includegraphics[width=0.98\linewidth]{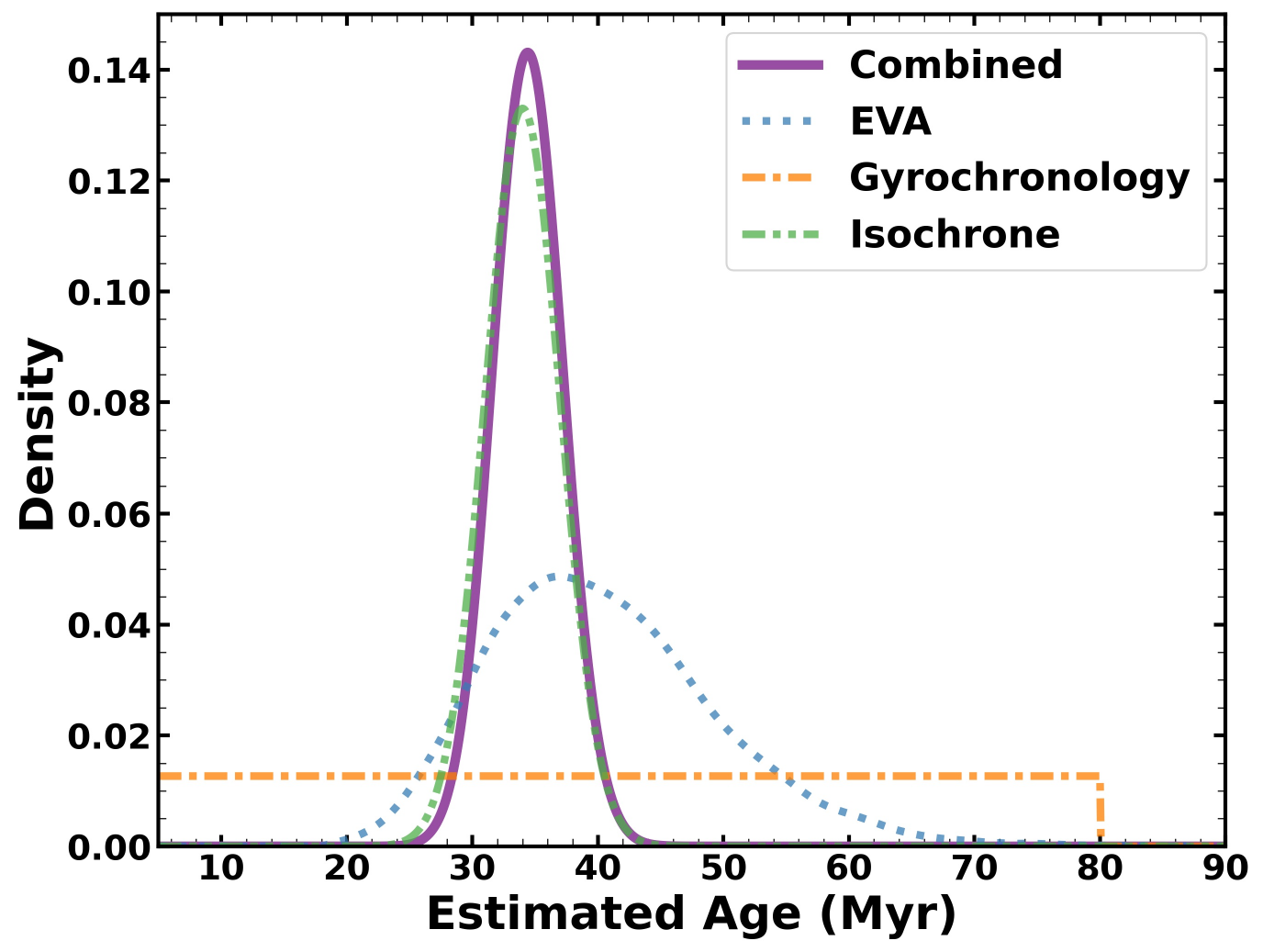}
    \caption{Final age estimate ($34\pm3$\,Myr; purple) compared to the age estimate from each method (colored lines). The final age is strongly influenced by the isochronal fit, which was far more precise due to the combination of high-mass evolving stars and pre-main-sequence M dwarfs. However, we highlight that all age metrics agree.}
    \label{fig:combine}
\end{figure}

We determine a final age for \starname\ by combining the individual age estimates (Figure \ref{fig:combine}). Using the likelihood distributions from each method, we find an overall age of $34\pm3$\,Myr for \starname\ and Vela Population IV. This age is in agreement with previous age determinations, though it provides the much tighter constraint necessary for planet evolution statistical modeling. 

\section{Discussion and Conclusions}
\label{sec:concl}

We report the validation of \planetname, a giant planet orbiting a young Sun-like star. We find \planetname\ to be $8.8\pm0.8$ R$_\oplus$ on a 14.844-day orbit. We derive the age of the host star, through the parent association, Vela Population IV, to be $34\pm3$\,Myr based on the CMD placements, rotation periods, and variability levels of the stars tightly co-moving with \starname.

\begin{figure}
    \centering
    \includegraphics[width=0.99\linewidth]{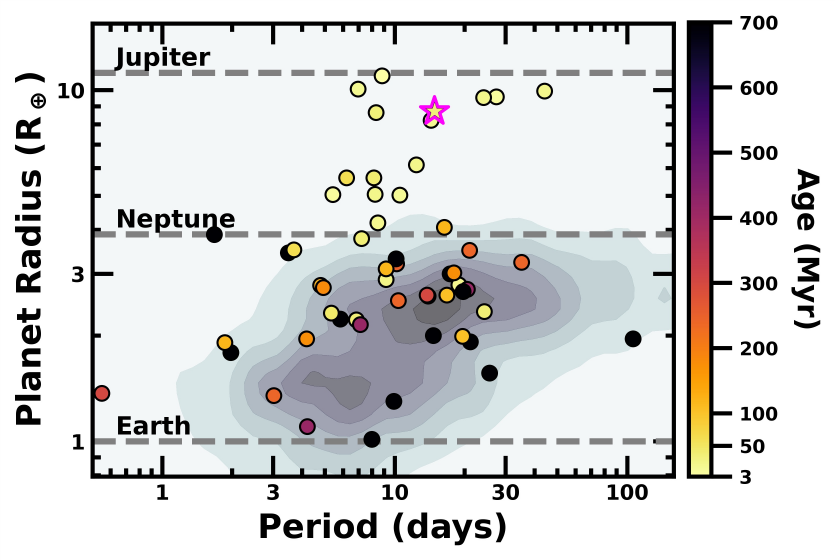}
    \caption{Distribution of planets discovered transiting members of young stellar associations (colored points) compared to the distribution of planets transiting (primarily) mature stars discovered using \kepler\ (background contours). TOI-6448\,b is shown as the star outlined in pink. The pale yellow points, representing the youngest population ($<$50\,Myr), tend to sit high in comparison to even slightly older 100--200\,Myr planets.}
    \label{fig:planet_pop}
\end{figure}

\planetname\ contributes to the growing population of $<$50\,Myr transiting planets that are ideal for understanding the evolutionary pathways that create the mature distribution of planets as discovered by \kepler. There are ten other $<$50\,Myr systems harboring 16 transiting planets. As can be seen in Figure~\ref{fig:planet_pop}, most of these land in the region of parameter space with few old planets from \kepler\ (5-11$R_\oplus$). This is consistent with the small number of mass measurements for these planets that indicate they are puffy sub-Neptunes or super-Earths \citep{Thao2024_featherweight,Barat2025}, and statistical surveys finding enhanced super-Neptune occurrence rates out to at least 200\,Myr \citep{Vach2024_ocr}.

A common concern with this conclusion is the role of observational bias; that the young planet population looks larger because high stellar variability is masking out the smaller planets. Observational bias can explain the deficit of $\lesssim2R_\oplus$ planets in the sample; injection/recovery tests on such stars suggest we are relatively insensitive to such small planets around $<50$\,Myr stars \citep[e.g.][]{Vach2024_ocr}. However, it cannot explain why we have detected {\it an excess} of $>5R_\oplus$ planets given the available sample of young stars. The most clear evidence that this is not bias is that stars from 80-200\,Myr are generally spinning {\it faster} than their younger counterparts as they are still spinning up from pre-main-sequence contraction. Indeed, some of the most challenging light curves for planet detection are $\simeq$100\,Myr stars \citep{Gaidos2017,Rizzuto2017}. This slightly older range also contains plenty of stars between Pleiades, $\alpha$ Per, Psc-Eri, and a number of Theia groups \citep{Kounkel2020}. Yet there are far fewer 5-11$R_\oplus$, 80-200\,Myr planets than there are similar-radius 0-50\,Myr planets. Indeed, this sharp drop in the occurrence of 5-11$R_\oplus$ past $\simeq$50\,Myr suggests the radii of these planets decrease rapidly, as suggested by a `gas dwarf' model \citep{Rogers2025}.

Counterintuitively, observational bias makes this planet size effect {\it stronger}. We can show this with a simple simulation. The THYME and TIDYE surveys have focused on just two regions for $<50$\,Myr stars: Taurus-Auriga and Sco-Cen, with \planetname{} as first in our search of Vela-Puppis. Of these three groups, there are 11 planets with periods $<$30\,days and radii of 5-11$R_\oplus$ that were discovered or recovered by these surveys; all found within a sample of 10,273 stars. If we assume the surveys are 100\% complete to planets in this radius and period range, and adopt a \kepler-like planet distribution \citep[star/planet properties from][]{Berger2020,Berger2023}, we should detect between 1 and 9 planets (95\%). For simplicity, this ignores differences in the target sample, and hence it is marginally consistent with our findings. If we assume completeness/sensitivity to these younger planets is $\simeq$80\% that of \kepler{}, the expected detections becomes 0--7 planets (95\%) and the difference becomes $>3\sigma$. The result is similarly significant if we count three candidates that have yet to be validated \citep[e.g.,][]{Vach2025_tic887} as real (even ignoring completeness).

As can be seen in Figure~\ref{fig:injrec}, 80\% completeness is likely generous, especially for planets from 15-30\,days where it is easy to confuse random stellar signals (flares and rotation) with planetary ones. Younger groups like Upper Scorpius and Taurus-Auriga also have a large population of dipper and burster variable stars \citep{Ansdell2016,Cody2017} for which our completeness is $\simeq$0. More importantly, {\it as we decrease sensitivity to young planets, the difference between young and old planets increases,} the opposite of our standard intuition. For the planet radii to be explained by observational bias one would need to assume that \kepler{} is missing the larger planets, despite a larger telescope, longer stare window, and better behaved stars.

\begin{figure}
    \centering
    \includegraphics[width=0.99\linewidth]{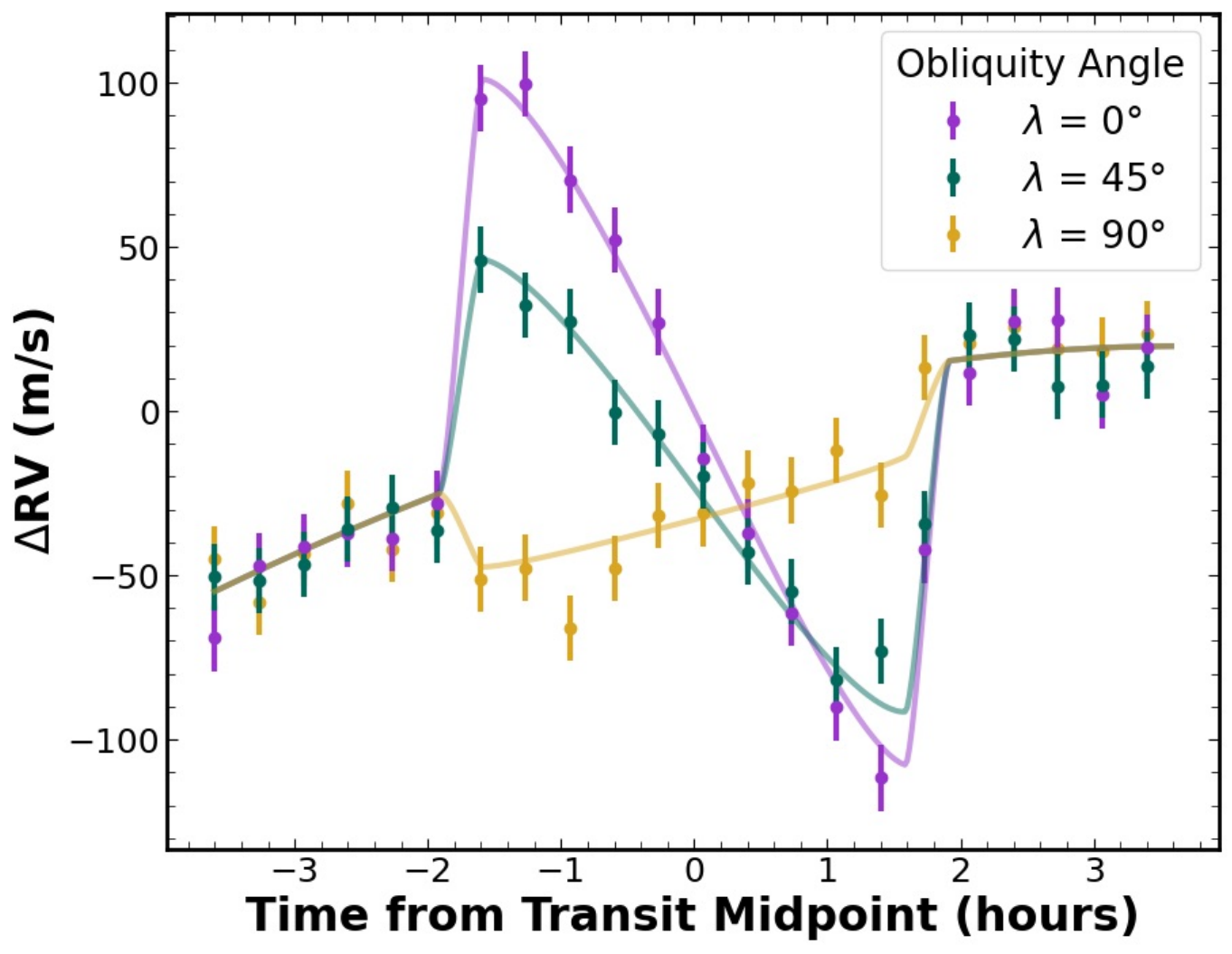}
    \caption{Simulated Rossiter-McLaughlin observations for \planetname\ for various sky-projected obliquity angles. Even at relatively long cadence (20-minutes), large uncertainties (10\,\ms), and considering stellar jitter ($\sim50$\,\ms\ over the transit duration), we can easily distinguish between aligned (purple) and misaligned (green and yellow) orbits. }
    \label{fig:rm_sim}
\end{figure}

Surveys using Rossiter-McLaughlin (RM) or Doppler tomography (DT) suggest that young planets are more aligned than their older counterparts \citep[e.g.][]{Zhou2020, Dai2020, Wirth2021, Johnson2022, Hirano2024}. However, the samples are still marginally consistent, primarily because the young planet sample is still too small. Young planets are {\it easier} targets, in part because of their larger radii, but also the rapid rotation rate of young stars \textit{increases} the RM amplitude. The stellar jitter is also increased, but this tends to work on the multi-day timescale of stellar rotation while RM and DT signals work on the $\sim$hours of the transit duration. As we show in Figure~\ref{fig:rm_sim}, the expected RM signal for \planetname{} is hundreds of \mps{} for an aligned system, while the jitter of the same time period is expected to be on the order of tens of \mps{}. 

Assuming this planet follows the trend of other young planets and predictions from models \citep{Rogers2025}, it will have a mass of 5-40$M_\oplus$, resulting in a radial velocity amplitude of just 2-8\mps. This is at least 1 order of magnitude below the expected stellar jitter ($\sim$100s of \mps based on the variability). However, prior studies have shown we can measure the masses of young puffy ($>5R_{\oplus}$) planets from their transmission spectra \citep[e.g.,][]{deWit2013, Thao2024_featherweight, Barat2025}. This method is {\it more} effective at lower masses, and works in the presence of strong spots. It is mostly limited by the availability of {\it JWST} time. 

\begin{figure}
    \centering
    \includegraphics[width=0.99\linewidth]{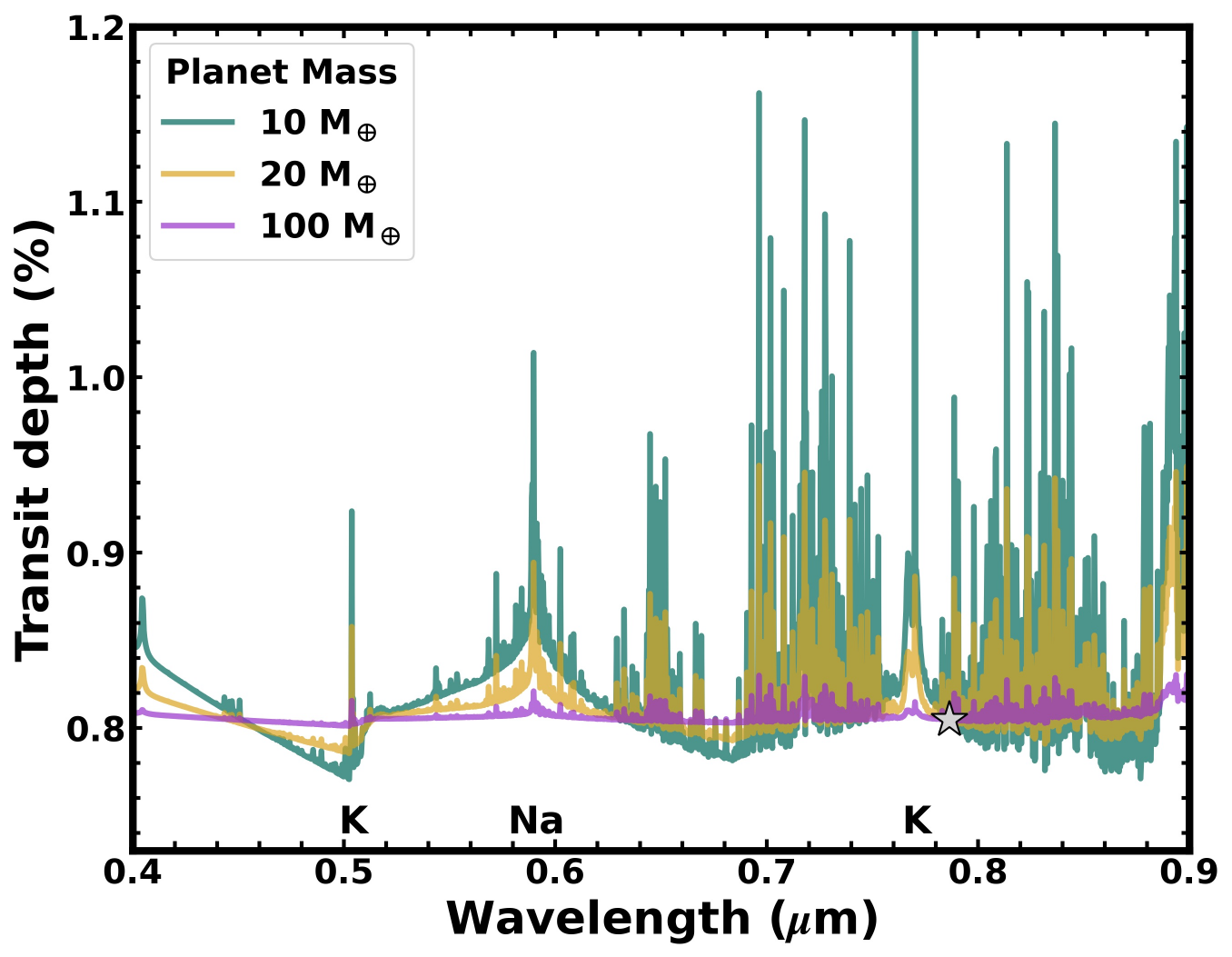}
    \caption{Simulated transmission spectra (R=15000) of \planetname\ for various planet masses. We label the major lines (Na and K) accessible from ground-based facilities. Each is normalized to the \tess\ depth at the \tess\ central wavelength (gray star). We can easily distinguish between a super-Earth progenitor (10$M_\oplus$, green), Neptune progenitor (20$M_\oplus$, yellow) and a Saturn planet (100$M_\oplus$, purple). }
    \label{fig:transmission_sim}
\end{figure}

However, rough masses for these kinds of systems might be possible from the ground. We model expected transmission spectra for \planetname\ assuming a super-Earth-progenitor mass (M $=10$M$_\oplus$), Neptune-progenitor mass (M $=20$M$_\oplus$) and a Saturn mass (M $=100$M$_\oplus$), using the basic parameters from \citet{Thao2024_featherweight} and PICASO \citep{Batalha2019}. We assume a cloudless atmosphere, consistent with the previous \jwst\ observations of HIP 67522\,b and V1298 Tau\,b. As we show in Figure~\ref{fig:transmission_sim}, these scenarios are distinguishable in the optical at a resolution of R$\simeq$15,000 and precision of a few mmag per resolving element; this is routinely achieved using high-precision spectrographs \citep[e.g.,][]{Benatti2021,Hirano2024}. Since H$_2$O bands are not accessible from the ground, the analysis relies on Na and K atomic lines in the optical, which are measurable due to the lower surface gravity. These features are also advantageous because the expected spot temperatures do not produce such lines \citep{Thao2023}.

\section*{Acknowledgments}

The authors would like to thank Halee and Bandit for their scientific feedback and reminders (demands) to take writing breaks. M.G.B. and A.W.M. were supported by NASA’s exoplanet research program (XRP 80NSSC25K7148), and M.G.B. was also supported by the NSF Graduate Research Fellowship (DGE-2040435). K.A.C. acknowledges support from the TESS mission via subaward s3449 from MIT. Funding for K.B. was provided by the European Union (ERC AdG SUBSTELLAR, GA 101054354). W.C.S. acknowledges funding from the NC Space Grant Undergraduate Research Scholarship and a Summer Undergraduate Research Fellowship from the Office for Undergraduate Research at the University of North Carolina at Chapel Hill.

Funding for the \tess\ mission is provided by NASA’s Science Mission Directorate. We acknowledge the use of public \tess\ data from pipelines at the \tess\ Science Office and at the \tess\ Science Processing Operations Center. Resources supporting this work were provided by the NASA High-End Computing (HEC) Program through the NASA Advanced Supercomputing (NAS) Division at Ames Research Center for the production of the SPOC data products. \tess\ data presented in this paper were obtained from the Mikulski Archive for Space Telescopes (MAST) at the Space Telescope Science Institute.

This research has made use of the Exoplanet Follow-up Observation Program (ExoFOP; DOI: 10.26134/ExoFOP5) website, which is operated by the California Institute of Technology, under contract with the National Aeronautics and Space Administration under the Exoplanet Exploration Program.

This research has used data from the CTIO/SMARTS 1.5m telescope, which is operated as part of the SMARTS Consortium by RECONS (www.recons.org) members Todd Henry, Hodari James, Wei-Chun Jao, and Leonardo Paredes.

This work has made use of data from the European Space Agency (ESA) mission Gaia (\url{https://www.cosmos.esa.int/gaia}), processed by the Gaia Data Processing and Analysis Consortium (DPAC; \url{https://www.cosmos.esa.int/web/gaia/dpac/consortium}). Funding for the DPAC has been provided by national institutions, in particular the institutions participating in the Gaia Multilateral Agreement.

This work makes use of observations from the LCOGT network. Part of the LCOGT telescope time was granted by NOIRLab through the Mid-Scale Innovations Program (MSIP). MSIP is funded by NSF.

This paper is based on observations made with the Las Cumbres Observatory’s education network telescopes that were upgraded through generous support from the Gordon and Betty Moore Foundation.

The research leading to these results has received funding from  the ARC grant for Concerted Research Actions, financed by the Wallonia-Brussels Federation. TRAPPIST is funded by the Belgian Fund for Scientific Research (Fond National de la Recherche Scientifique, FNRS) under the grant PDR T.0120.21. M.G. is F.R.S.-FNRS Research Director.

\bibliography{planetSearch,mannbib}

\begin{thebibliography}{}
\expandafter\ifx\csname natexlab\endcsname\relax\def\natexlab#1{#1}\fi
\providecommand{\url}[1]{\href{#1}{#1}}
\providecommand{\dodoi}[1]{doi:~\href{http://doi.org/#1}{\nolinkurl{#1}}}
\providecommand{\doeprint}[1]{\href{http://ascl.net/#1}{\nolinkurl{http://ascl.net/#1}}}
\providecommand{\doarXiv}[1]{\href{https://arxiv.org/abs/#1}{\nolinkurl{https://arxiv.org/abs/#1}}}

\bibitem[{{Allard} {et~al.}(2013){Allard}, {Homeier}, {Freytag}, {Schaffenberger}, {}, \& {Rajpurohit}}]{Allard2013}
{Allard}, F., {Homeier}, D., {Freytag}, B., {et~al.} 2013, Memorie della Societa Astronomica Italiana Supplementi, 24, 128.
\newblock \doarXiv{1302.6559}

\bibitem[{{Alves} {et~al.}(2025){Alves}, {Jenkins}, {Vines}, {Battley}, {Lendl}, {Bouchy}, {Nielsen}, {Gill}, {Moyano}, {Anderson}, {Burleigh}, {Casewell}, {Goad}, {Hawthorn}, {Kendall}, {McCormac}, {Osborn}, {Smith}, {Udry}, {Wheatley}, {Saha}, {Parc}, {Nigioni}, {Apergis}, \& {Ramsay}}]{Alves2025_ngst33}
{Alves}, D.~R., {Jenkins}, J.~S., {Vines}, J.~I., {et~al.} 2025, \mnras, 536, 1538, \dodoi{10.1093/mnras/stae2582}

\bibitem[{{Ansdell} {et~al.}(2016){Ansdell}, {Gaidos}, {Rappaport}, {Jacobs}, {LaCourse}, {Jek}, {Mann}, {Wyatt}, {Kennedy}, {Williams}, \& {Boyajian}}]{Ansdell2016}
{Ansdell}, M., {Gaidos}, E., {Rappaport}, S.~A., {et~al.} 2016, \apj, 816, 69, \dodoi{10.3847/0004-637X/816/2/69}

\bibitem[{Barat {et~al.}(2024)Barat, Désert, Goyal, Vazan, Kawashima, Fortney, Bean, Line, Panwar, Jacobs, Shivkumar, Sikora, Baeyens, Oklopcić, David, \& Livingston}]{Barat2024}
Barat, S., Désert, J.-M., Goyal, J.~M., {et~al.} 2024, First Comparative Exoplanetology Within a Transiting Multi-planet System: Comparing the atmospheres of V1298 Tau b and c.
\newblock \doarXiv{2407.14995}

\bibitem[{{Barat} {et~al.}(2025){Barat}, {D{\'e}sert}, {Mukherjee}, {Goyal}, {Xue}, {Kawashima}, {Vazan}, {Misener}, {Schlichting}, {Fortney}, {Bean}, {Avarsekar}, {Henry}, {Baeyens}, {Line}, {Livingston}, {David}, {Petigura}, {Sikora}, {Shivkumar}, {Feinstein}, \& {Oklop{\v{c}}i{\'c}}}]{Barat2025}
{Barat}, S., {D{\'e}sert}, J.-M., {Mukherjee}, S., {et~al.} 2025, arXiv e-prints, arXiv:2507.08837, \dodoi{10.48550/arXiv.2507.08837}

\bibitem[{Barber \& Mann(2023)}]{BarberMann2023}
Barber, M.~G., \& Mann, A.~W. 2023, The Astrophysical Journal, 953, 127, \dodoi{10.3847/1538-4357/ace044}

\bibitem[{{Barber} {et~al.}(2025){Barber}, {Mann}, {Vanderburg}, {Boyle}, \& {Lopez Murillo}}]{Barber2025_2076e}
{Barber}, M.~G., {Mann}, A.~W., {Vanderburg}, A., {Boyle}, A.~W., \& {Lopez Murillo}, A.~I. 2025, arXiv e-prints, arXiv:2505.06358, \dodoi{10.48550/arXiv.2505.06358}

\bibitem[{{Barber} {et~al.}(2024{\natexlab{a}}){Barber}, {Thao}, {Mann}, {Vanderburg}, {Mori}, {Livingston}, {Fukui}, {Narita}, {Kraus}, {Tofflemire}, {Newton}, {Winn}, {Jenkins}, {Seager}, {Collins}, \& {Twicken}}]{Barber2024_hipc}
{Barber}, M.~G., {Thao}, P.~C., {Mann}, A.~W., {et~al.} 2024{\natexlab{a}}, \apjl, 973, L30, \dodoi{10.3847/2041-8213/ad77d9}

\bibitem[{{Barber} {et~al.}(2024{\natexlab{b}}){Barber}, {Mann}, {Vanderburg}, {Krolikowski}, {Kraus}, {Ansdell}, {Pearce}, {Mace}, {Andrews}, {Boyle}, {Collins}, {De Furio}, {Dragomir}, {Espaillat}, {Feinstein}, {Fields}, {Jaffe}, {Lopez Murillo}, {Murgas}, {Newton}, {Palle}, {Sawczynec}, {Schwarz}, {Thao}, {Tofflemire}, {Watkins}, {Jenkins}, {Latham}, {Ricker}, {Seager}, {Vanderspek}, {Winn}, {Charbonneau}, {Essack}, {Rodriguez}, {Shporer}, {Twicken}, \& {Villase{\~n}or}}]{Barber2024_iras}
{Barber}, M.~G., {Mann}, A.~W., {Vanderburg}, A., {et~al.} 2024{\natexlab{b}}, \nat, 635, 574, \dodoi{10.1038/s41586-024-08123-3}

\bibitem[{{Batalha} {et~al.}(2019){Batalha}, {Marley}, {Lewis}, \& {Fortney}}]{Batalha2019}
{Batalha}, N.~E., {Marley}, M.~S., {Lewis}, N.~K., \& {Fortney}, J.~J. 2019, \apj, 878, 70, \dodoi{10.3847/1538-4357/ab1b51}

\bibitem[{Benatti {et~al.}(2021)Benatti, Damasso, Borsa, Locci, Pillitteri, Desidera, Maggio, Micela, Wolk, Claudi, Malavolta, \& Modirrousta-Galian}]{Benatti2021}
Benatti, S., Damasso, M., Borsa, F., {et~al.} 2021, Astronomy \& Astrophysics, 650, A66, \dodoi{10.1051/0004-6361/202140416}

\bibitem[{{Berger} {et~al.}(2020){Berger}, {Huber}, {Gaidos}, {van Saders}, \& {Weiss}}]{Berger2020}
{Berger}, T.~A., {Huber}, D., {Gaidos}, E., {van Saders}, J.~L., \& {Weiss}, L.~M. 2020, \aj, 160, 108, \dodoi{10.3847/1538-3881/aba18a}

\bibitem[{{Berger} {et~al.}(2023){Berger}, {Schlieder}, \& {Huber}}]{Berger2023}
{Berger}, T.~A., {Schlieder}, J.~E., \& {Huber}, D. 2023, arXiv e-prints, arXiv:2301.11338, \dodoi{10.48550/arXiv.2301.11338}

\bibitem[{Bernstein {et~al.}(2003)Bernstein, Shectman, Gunnels, Mochnacki, \& Athey}]{Bernstein2003}
Bernstein, R., Shectman, S.~A., Gunnels, S.~M., Mochnacki, S., \& Athey, A.~E. 2003, in Instrument Design and Performance for Optical/Infrared Ground-based Telescopes, ed. M.~Iye \& A.~F.~M. Moorwood, Vol. 4841, International Society for Optics and Photonics (SPIE), 1694 -- 1704, \dodoi{10.1117/12.461502}

\bibitem[{{Blackwell} \& {Shallis}(1977)}]{Blackwell1977}
{Blackwell}, D.~E., \& {Shallis}, M.~J. 1977, \mnras, 180, 177, \dodoi{10.1093/mnras/180.2.177}

\bibitem[{{Blunt} {et~al.}(2023){Blunt}, {Carvalho}, {David}, {Beichman}, {Zink}, {Gaidos}, {Behmard}, {Bouma}, {Cody}, {Dai}, {Foreman-Mackey}, {Grunblatt}, {Howard}, {Kosiarek}, {Knutson}, {Rubenzahl}, {Beard}, {Chontos}, {Giacalone}, {Hirano}, {Johnson}, {Lubin}, {Akana Murphy}, {Petigura}, {Van Zandt}, \& {Weiss}}]{Blunt2023}
{Blunt}, S., {Carvalho}, A., {David}, T.~J., {et~al.} 2023, \aj, 166, 62, \dodoi{10.3847/1538-3881/acde78}

\bibitem[{{Bossini, D.} {et~al.}(2019){Bossini, D.}, {Vallenari, A.}, {Bragaglia, A.}, {Cantat-Gaudin, T.}, {Sordo, R.}, {Balaguer-Núñez, L.}, {Jordi, C.}, {Moitinho, A.}, {Soubiran, C.}, {Casamiquela, L.}, {Carrera, R.}, \& {Heiter, U.}}]{Bossini2019}
{Bossini, D.}, {Vallenari, A.}, {Bragaglia, A.}, {et~al.} 2019, A\&A, 623, A108, \dodoi{10.1051/0004-6361/201834693}

\bibitem[{{Bouma} {et~al.}(2023){Bouma}, {Palumbo}, \& {Hillenbrand}}]{Bouma2023}
{Bouma}, L.~G., {Palumbo}, E.~K., \& {Hillenbrand}, L.~A. 2023, \apjl, 947, L3, \dodoi{10.3847/2041-8213/acc589}

\bibitem[{{Bouma} {et~al.}(2022){Bouma}, {Kerr}, {Curtis}, {Isaacson}, {Hillenbrand}, {Howard}, {Kraus}, {Bieryla}, {Latham}, {Petigura}, \& {Huber}}]{Bouma2022}
{Bouma}, L.~G., {Kerr}, R., {Curtis}, J.~L., {et~al.} 2022, \aj, 164, 215, \dodoi{10.3847/1538-3881/ac93ff}

\bibitem[{{Boyle} {et~al.}(2025){Boyle}, {Mann}, \& {Bush}}]{Boyle2025}
{Boyle}, A.~W., {Mann}, A.~W., \& {Bush}, J. 2025, arXiv e-prints, arXiv:2504.13262.
\newblock \doarXiv{2504.13262}

\bibitem[{Bressan {et~al.}(2012)Bressan, Marigo, Girardi, Salasnich, Dal~Cero, Rubele, \& Nanni}]{Bressan2012}
Bressan, A., Marigo, P., Girardi, L., {et~al.} 2012, Monthly Notices of the Royal Astronomical Society, 427, 127, \dodoi{10.1111/j.1365-2966.2012.21948.x}

\bibitem[{{Brown} {et~al.}(2013){Brown}, {Baliber}, {Bianco}, {Bowman}, {Burleson}, {Conway}, {Crellin}, {Depagne}, {De Vera}, {Dilday}, {Dragomir}, {Dubberley}, {Eastman}, {Elphick}, {Falarski}, {Foale}, {Ford}, {Fulton}, {Garza}, {Gomez}, {Graham}, {Greene}, {Haldeman}, {Hawkins}, {Haworth}, {Haynes}, {Hidas}, {Hjelstrom}, {Howell}, {Hygelund}, {Lister}, {Lobdill}, {Martinez}, {Mullins}, {Norbury}, {Parrent}, {Paulson}, {Petry}, {Pickles}, {Posner}, {Rosing}, {Ross}, {Sand}, {Saunders}, {Shobbrook}, {Shporer}, {Street}, {Thomas}, {Tsapras}, {Tufts}, {Valenti}, {Vander Horst}, {Walker}, {White}, \& {Willis}}]{Brown:2013}
{Brown}, T.~M., {Baliber}, N., {Bianco}, F.~B., {et~al.} 2013, \pasp, 125, 1031, \dodoi{10.1086/673168}

\bibitem[{{Burn} {et~al.}(2024){Burn}, {Mordasini}, {Mishra}, {Haldemann}, {Venturini}, {Emsenhuber}, \& {Henning}}]{Burn2024}
{Burn}, R., {Mordasini}, C., {Mishra}, L., {et~al.} 2024, Nature Astronomy, 8, 463, \dodoi{10.1038/s41550-023-02183-7}

\bibitem[{{Cantat-Gaudin, T.} {et~al.}(2019){Cantat-Gaudin, T.}, {Jordi, C.}, {Wright, N. J.}, {Armstrong, J. J.}, {Vallenari, A.}, {Balaguer-Núñez, L.}, {Ramos, P.}, {Bossini, D.}, {Padoan, P.}, {Pelkonen, V. M.}, {Mapelli, M.}, \& {Jeffries, R. D.}}]{CantatGaudin2019}
{Cantat-Gaudin, T.}, {Jordi, C.}, {Wright, N. J.}, {et~al.} 2019, A\&A, 626, A17, \dodoi{10.1051/0004-6361/201834957}

\bibitem[{{Capistrant} {et~al.}(2024){Capistrant}, {Soares-Furtado}, {Vanderburg}, {Jankowski}, {Mann}, {Ross}, {Srdoc}, {Hinkel}, {Becker}, {Magliano}, {Limbach}, {Stephan}, {Nine}, {Tofflemire}, {Kraus}, {Giacalone}, {Winn}, {Bieryla}, {Bouma}, {Ciardi}, {Collins}, {Covone}, {de Beurs}, {Huang}, {Jenkins}, {Kreidberg}, {Latham}, {Quinn}, {Seager}, {Shporer}, {Twicken}, {Wohler}, {Vanderspek}, {Yarza}, \& {Ziegler}}]{Capistrant2024}
{Capistrant}, B.~K., {Soares-Furtado}, M., {Vanderburg}, A., {et~al.} 2024, \aj, 167, 54, \dodoi{10.3847/1538-3881/ad1039}

\bibitem[{{Cody} {et~al.}(2017){Cody}, {Hillenbrand}, {David}, {Carpenter}, {Everett}, \& {Howell}}]{Cody2017}
{Cody}, A.~M., {Hillenbrand}, L.~A., {David}, T.~J., {et~al.} 2017, \apj, 836, 41, \dodoi{10.3847/1538-4357/836/1/41}

\bibitem[{{Collins}(2019)}]{collins:2019}
{Collins}, K. 2019, in American Astronomical Society Meeting Abstracts, Vol. 233, American Astronomical Society Meeting Abstracts \#233, 140.05

\bibitem[{{Collins} {et~al.}(2017){Collins}, {Kielkopf}, {Stassun}, \& {Hessman}}]{Collins:2017}
{Collins}, K.~A., {Kielkopf}, J.~F., {Stassun}, K.~G., \& {Hessman}, F.~V. 2017, \aj, 153, 77, \dodoi{10.3847/1538-3881/153/2/77}

\bibitem[{{Cutri} \& {et al.}(2014)}]{allwise}
{Cutri}, R.~M., \& {et al.} 2014, VizieR Online Data Catalog, II/328

\bibitem[{{Dai} {et~al.}(2020){Dai}, {Roy}, {Fulton}, {Robertson}, {Hirsch}, {Isaacson}, {Albrecht}, {Mann}, {Kristiansen}, {Batalha}, {Beard}, {Behmard}, {Chontos}, {Crossfield}, {Dalba}, {Dressing}, {Giacalone}, {Hill}, {Howard}, {Huber}, {Kane}, {Kosiarek}, {Lubin}, {Mayo}, {Mocnik}, {Akana Murphy}, {Petigura}, {Rosenthal}, {Rubenzahl}, {Scarsdale}, {Weiss}, {Van Zandt}, {Ricker}, {Vanderspek}, {Latham}, {Seager}, {Winn}, {Jenkins}, {Caldwell}, {Charbonneau}, {Daylan}, {G{\"u}nther}, {Morgan}, {Quinn}, {Rose}, \& {Smith}}]{Dai2020}
{Dai}, F., {Roy}, A., {Fulton}, B., {et~al.} 2020, \aj, 160, 193, \dodoi{10.3847/1538-3881/abb3bd}

\bibitem[{{Dai} {et~al.}(2024){Dai}, {Goldberg}, {Batygin}, {van Saders}, {Chiang}, {Choksi}, {Li}, {Petigura}, {Gilbert}, {Millholland}, {Dai}, {Bouma}, {Weiss}, \& {Winn}}]{Dai2024}
{Dai}, F., {Goldberg}, M., {Batygin}, K., {et~al.} 2024, \aj, 168, 239, \dodoi{10.3847/1538-3881/ad83a6}

\bibitem[{{David} {et~al.}(2019{\natexlab{a}}){David}, {Petigura}, {Luger}, {Foreman-Mackey}, {Livingston}, {Mamajek}, \& {Hillenbrand}}]{David2019b_v1298}
{David}, T.~J., {Petigura}, E.~A., {Luger}, R., {et~al.} 2019{\natexlab{a}}, \apjl, 885, L12, \dodoi{10.3847/2041-8213/ab4c99}

\bibitem[{{David} {et~al.}(2019{\natexlab{b}}){David}, {Cody}, {Hedges}, {Mamajek}, {Hillenbrand}, {Ciardi}, {Beichman}, {Petigura}, {Fulton}, {Isaacson}, {Howard}, {Gagn{\'e}}, {Saunders}, {Rebull}, {Stauffer}, {Vasisht}, \& {Hinkley}}]{David2019a_v1298}
{David}, T.~J., {Cody}, A.~M., {Hedges}, C.~L., {et~al.} 2019{\natexlab{b}}, \aj, 158, 79, \dodoi{10.3847/1538-3881/ab290f}

\bibitem[{{de Wit} \& {Seager}(2013)}]{deWit2013}
{de Wit}, J., \& {Seager}, S. 2013, Science, 342, 1473, \dodoi{10.1126/science.1245450}

\bibitem[{{Dias, W. S.} {et~al.}(2002){Dias, W. S.}, {Alessi, B. S.}, {Moitinho, A.}, \& {Lépine, J. R. D.}}]{Dias2002}
{Dias, W. S.}, {Alessi, B. S.}, {Moitinho, A.}, \& {Lépine, J. R. D.} 2002, A\&A, 389, 871, \dodoi{10.1051/0004-6361:20020668}

\bibitem[{{Dotter} {et~al.}(2008){Dotter}, {Chaboyer}, {Jevremovi{\'c}}, {Kostov}, {Baron}, \& {Ferguson}}]{Dotter2008}
{Dotter}, A., {Chaboyer}, B., {Jevremovi{\'c}}, D., {et~al.} 2008, \apjs, 178, 89, \dodoi{10.1086/589654}

\bibitem[{{Evans} {et~al.}(2018){Evans}, {Riello}, {De Angeli}, {Carrasco}, {Montegriffo}, {Fabricius}, {Jordi}, {Palaversa}, {Diener}, {Busso}, {Cacciari}, {van Leeuwen}, {Burgess}, {Davidson}, {Harrison}, {Hodgkin}, {Pancino}, {Richards}, {Altavilla}, {Balaguer-N{\'u}{\~n}ez}, {Barstow}, {Bellazzini}, {Brown}, {Castellani}, {Cocozza}, {De Luise}, {Delgado}, {Ducourant}, {Galleti}, {Gilmore}, {Giuffrida}, {Holl}, {Kewley}, {Koposov}, {Marinoni}, {Marrese}, {Osborne}, {Piersimoni}, {Portell}, {Pulone}, {Ragaini}, {Sanna}, {Terrett}, {Walton}, {Wevers}, \& {Wyrzykowski}}]{Evans2018}
{Evans}, D.~W., {Riello}, M., {De Angeli}, F., {et~al.} 2018, \aap, 616, A4, \dodoi{10.1051/0004-6361/201832756}

\bibitem[{{Feiden}(2016)}]{Feiden2016}
{Feiden}, G.~A. 2016, \aap, 593, A99, \dodoi{10.1051/0004-6361/201527613}

\bibitem[{{Fernandes} {et~al.}(2022){Fernandes}, {Mulders}, {Pascucci}, {Bergsten}, {Koskinen}, {Hardegree-Ullman}, {Pearson}, {Giacalone}, {Zink}, {Ciardi}, \& {O'Brien}}]{Fernandes2022}
{Fernandes}, R.~B., {Mulders}, G.~D., {Pascucci}, I., {et~al.} 2022, \aj, 164, 78, \dodoi{10.3847/1538-3881/ac7b29}

\bibitem[{Fields {et~al.}(2025)Fields, Mann, Kesseli, \& Boyle}]{Fields2025}
Fields, M.~J., Mann, A.~W., Kesseli, A., \& Boyle, A.~W. 2025, Disk-Star Alignment I: Pre-Main-Sequence Stellar Parameters and the Statistical Alignment Between Disks and Stellar Rotation.
\newblock \doarXiv{2504.02990}

\bibitem[{{Foreman-Mackey}(2018)}]{celerite2}
{Foreman-Mackey}, D. 2018, Research Notes of the American Astronomical Society, 2, 31, \dodoi{10.3847/2515-5172/aaaf6c}

\bibitem[{Foreman-Mackey {et~al.}(2017)Foreman-Mackey, Agol, Ambikasaran, \& Angus}]{ForemanMackey2017}
Foreman-Mackey, D., Agol, E., Ambikasaran, S., \& Angus, R. 2017, The Astronomical Journal, 154, 220, \dodoi{10.3847/1538-3881/aa9332}

\bibitem[{{Foreman-Mackey} {et~al.}(2013){Foreman-Mackey}, {Hogg}, {Lang}, \& {Goodman}}]{emcee}
{Foreman-Mackey}, D., {Hogg}, D.~W., {Lang}, D., \& {Goodman}, J. 2013, \pasp, 125, 306, \dodoi{10.1086/670067}

\bibitem[{{Gagn{\'e}} {et~al.}(2020){Gagn{\'e}}, {David}, {Mamajek}, {Mann}, {Faherty}, \& {B{\'e}dard}}]{Gagne2020}
{Gagn{\'e}}, J., {David}, T.~J., {Mamajek}, E.~E., {et~al.} 2020, \apj, 903, 96, \dodoi{10.3847/1538-4357/abb77e}

\bibitem[{{Gaia Collaboration} {et~al.}(2023){Gaia Collaboration}, {Vallenari}, {Brown}, {Prusti}, {de Bruijne}, {Arenou}, {Babusiaux}, {Biermann}, {Creevey}, {Ducourant}, \& et~al.}]{GaiaCollaboration2023}
{Gaia Collaboration}, {Vallenari}, A., {Brown}, A.~G.~A., {et~al.} 2023, \aap, 674, A1, \dodoi{10.1051/0004-6361/202243940}

\bibitem[{{Gaidos} {et~al.}(2014){Gaidos}, {Mann}, {L{\'e}pine}, {Buccino}, {James}, {Ansdell}, {Petrucci}, {Mauas}, \& {Hilton}}]{Gaidos2014}
{Gaidos}, E., {Mann}, A.~W., {L{\'e}pine}, S., {et~al.} 2014, \mnras, 443, 2561, \dodoi{10.1093/mnras/stu1313}

\bibitem[{{Gaidos} {et~al.}(2017){Gaidos}, {Mann}, {Rizzuto}, {Nofi}, {Mace}, {Vanderburg}, {Feiden}, {Narita}, {Takeda}, {Esposito}, {De Rosa}, {Ansdell}, {Hirano}, {Graham}, {Kraus}, \& {Jaffe}}]{Gaidos2017}
{Gaidos}, E., {Mann}, A.~W., {Rizzuto}, A., {et~al.} 2017, \mnras, 464, 850, \dodoi{10.1093/mnras/stw2345}

\bibitem[{{Giacalone} \& {Dressing}(2020)}]{triceratops}
{Giacalone}, S., \& {Dressing}, C.~D. 2020, {triceratops: Candidate exoplanet rating tool}.
\newblock \doeprint{2002.004}

\bibitem[{Gilbert {et~al.}(2022)Gilbert, Barclay, Quintana, Walkowicz, Vega, Schlieder, Monsue, Cale, Collins, Gaidos, Mufti, Reefe, Plavchan, Tanner, Wittenmyer, Wittrock, Jenkins, Latham, Ricker, Rose, Seager, Vanderspek, \& Winn}]{Gilbert2022}
Gilbert, E.~A., Barclay, T., Quintana, E.~V., {et~al.} 2022, The Astronomical Journal, 163, 147, \dodoi{10.3847/1538-3881/ac23ca}

\bibitem[{Gillon {et~al.}(2011)Gillon, Jehin, Magain, Chantry, Hutsem{\'{e}}kers, Manfroid, Queloz, \& Udry}]{Gillon2011}
Gillon, M., Jehin, E., Magain, P., {et~al.} 2011, {EPJ} Web of Conferences, 11, 06002, \dodoi{10.1051/epjconf/20101106002}

\bibitem[{{Ginzburg} {et~al.}(2016){Ginzburg}, {Schlichting}, \& {Sari}}]{Ginzburg2016}
{Ginzburg}, S., {Schlichting}, H.~E., \& {Sari}, R. 2016, \apj, 825, 29, \dodoi{10.3847/0004-637X/825/1/29}

\bibitem[{Goodman \& Weare(2010)}]{goodman2010}
Goodman, J., \& Weare, J. 2010, Commun. Appl. Math. Comput. Sci., 5, 65, \dodoi{10.2140/camcos.2010.5.65}

\bibitem[{{Hattori} {et~al.}(2022){Hattori}, {Foreman-Mackey}, {Hogg}, {Montet}, {Angus}, {Pritchard}, {Curtis}, \& {Sch{\"o}lkopf}}]{2022AJ....163..284H}
{Hattori}, S., {Foreman-Mackey}, D., {Hogg}, D.~W., {et~al.} 2022, \aj, 163, 284, \dodoi{10.3847/1538-3881/ac625a}

\bibitem[{{Heap} \& {Lindler}(2007)}]{Heap2007}
{Heap}, S.~R., \& {Lindler}, D.~J. 2007, in Astronomical Society of the Pacific Conference Series, Vol. 374, From Stars to Galaxies: Building the Pieces to Build Up the Universe, ed. A.~{Vallenari}, R.~{Tantalo}, L.~{Portinari}, \& A.~{Moretti}, 409

\bibitem[{{Henden} {et~al.}(2012){Henden}, {Levine}, {Terrell}, {Smith}, \& {Welch}}]{Henden2012}
{Henden}, A.~A., {Levine}, S.~E., {Terrell}, D., {Smith}, T.~C., \& {Welch}, D. 2012, Journal of the American Association of Variable Star Observers (JAAVSO), 40, 430

\bibitem[{Hirano {et~al.}(2024)Hirano, Gaidos, Harakawa, Hodapp, Kotani, Kudo, Kurokawa, Kuzuhara, Mann, Nishikawa, Omiya, Serizawa, Tamura, Thao, Ueda, \& Vievard}]{Hirano2024}
Hirano, T., Gaidos, E., Harakawa, H., {et~al.} 2024, Monthly Notices of the Royal Astronomical Society, 530, 3117, \dodoi{10.1093/mnras/stae998}

\bibitem[{Husser {et~al.}(2013)Husser, Wende-von Berg, Dreizler, Homeier, Reiners, Barman, \& Hauschildt}]{Husser2013}
Husser, T.-O., Wende-von Berg, S., Dreizler, S., {et~al.} 2013, Astronomy \& Astrophysics, 553, A6, \dodoi{10.1051/0004-6361/201219058}

\bibitem[{{Jeffries} {et~al.}(2023){Jeffries}, {Jackson}, {Wright}, {Weaver}, {Gilmore}, {Randich}, {Bragaglia}, {Korn}, {Smiljanic}, {Biazzo}, {Casey}, {Frasca}, {Gonneau}, {Guiglion}, {Morbidelli}, {Prisinzano}, {Sacco}, {Tautvai{\v{s}}ien{\.{e}}}, {Worley}, \& {Zaggia}}]{Jeffries2023}
{Jeffries}, R.~D., {Jackson}, R.~J., {Wright}, N.~J., {et~al.} 2023, \mnras, 523, 802, \dodoi{10.1093/mnras/stad1293}

\bibitem[{{Jehin} {et~al.}(2011){Jehin}, {Gillon}, {Queloz}, {Magain}, {Manfroid}, {Chantry}, {Lendl}, {Hutsem{\'e}kers}, \& {Udry}}]{Jehin2011}
{Jehin}, E., {Gillon}, M., {Queloz}, D., {et~al.} 2011, The Messenger, 145, 2

\bibitem[{{Johnson} {et~al.}(2018){Johnson}, {Dai}, {Justesen}, {Gandolfi}, {Hatzes}, {Nowak}, {Endl}, {Cochran}, {Hidalgo}, {Watanabe}, {Parviainen}, {Hirano}, {Villanueva}, {Prieto-Arranz}, {Narita}, {Palle}, {Guenther}, {Barrag{\'a}n}, {Trifonov}, {Niraula}, {MacQueen}, {Cabrera}, {Csizmadia}, {Eigm{\"u}ller}, {Grziwa}, {Korth}, {P{\"a}tzold}, {Smith}, {Albrecht}, {Alonso}, {Deeg}, {Erikson}, {Esposito}, {Fridlund}, {Fukui}, {Kusakabe}, {Kuzuhara}, {Livingston}, {Monta{\~n}es Rodriguez}, {Nespral}, {Persson}, {Purismo}, {Raimundo}, {Rauer}, {Ribas}, {Tamura}, {Van Eylen}, \& {Winn}}]{MISTTBORN}
{Johnson}, M.~C., {Dai}, F., {Justesen}, A.~B., {et~al.} 2018, \mnras, 481, 596, \dodoi{10.1093/mnras/sty2238}

\bibitem[{{Johnson} {et~al.}(2022){Johnson}, {David}, {Petigura}, {Isaacson}, {Van Zandt}, {Ilyin}, {Strassmeier}, {Mallonn}, {Zhou}, {Mann}, {Livingston}, {Luger}, {Dai}, {Weiss}, {Mo{\v{c}}nik}, {Giacalone}, {Hill}, {Rice}, {Blunt}, {Rubenzahl}, {Dalba}, {Esquerdo}, {Berlind}, {Calkins}, \& {Foreman-Mackey}}]{Johnson2022}
{Johnson}, M.~C., {David}, T.~J., {Petigura}, E.~A., {et~al.} 2022, \aj, 163, 247, \dodoi{10.3847/1538-3881/ac6271}

\bibitem[{{Karalis} {et~al.}(2025){Karalis}, {Lee}, \& {Thorngren}}]{Karalis2025}
{Karalis}, A., {Lee}, E.~J., \& {Thorngren}, D.~P. 2025, \apj, 978, 46, \dodoi{10.3847/1538-4357/ad946c}

\bibitem[{{Katz} {et~al.}(2019){Katz}, {Sartoretti}, {Cropper}, {Panuzzo}, {Seabroke}, {Viala}, {Benson}, {Blomme}, {Jasniewicz}, {Jean-Antoine}, {Huckle}, {Smith}, {Baker}, {Crifo}, {Damerdji}, {David}, {Dolding}, {Fr{\'e}mat}, {Gosset}, {Guerrier}, {Guy}, {Haigron}, {Jan{\ss}en}, {Marchal}, {Plum}, {Soubiran}, {Th{\'e}venin}, {Ajaj}, {Allende Prieto}, {Babusiaux}, {Boudreault}, {Chemin}, {Delle Luche}, {Fabre}, {Gueguen}, {Hambly}, {Lasne}, {Meynadier}, {Pailler}, {Panem}, {Royer}, {Tauran}, {Zurbach}, {Zwitter}, {Arenou}, {Bossini}, {Gerssen}, {G{\'o}mez}, {Lemaitre}, {Leclerc}, {Morel}, {Munari}, {Turon}, {Vallenari}, \& {{\v{Z}}erjal}}]{Katz2019}
{Katz}, D., {Sartoretti}, P., {Cropper}, M., {et~al.} 2019, \aap, 622, A205, \dodoi{10.1051/0004-6361/201833273}

\bibitem[{Kelson(2003)}]{Kelson2003}
Kelson, D.~D. 2003, Publications of the Astronomical Society of the Pacific, 115, 688, \dodoi{10.1086/375502}

\bibitem[{Kelson {et~al.}(2000)Kelson, Illingworth, van Dokkum, \& Franx}]{Kelson2000}
Kelson, D.~D., Illingworth, G.~D., van Dokkum, P.~G., \& Franx, M. 2000, The Astrophysical Journal, 531, 159, \dodoi{10.1086/308445}

\bibitem[{{Kenyon} \& {Hartmann}(1995)}]{Kenyon1995}
{Kenyon}, S.~J., \& {Hartmann}, L. 1995, \apjs, 101, 117, \dodoi{10.1086/192235}

\bibitem[{{Kesseli} {et~al.}(2018){Kesseli}, {Muirhead}, {Mann}, \& {Mace}}]{Kesseli2018}
{Kesseli}, A.~Y., {Muirhead}, P.~S., {Mann}, A.~W., \& {Mace}, G. 2018, \aj, 155, 225, \dodoi{10.3847/1538-3881/aabccb}

\bibitem[{{Kharchenko, N. V.} {et~al.}(2013){Kharchenko, N. V.}, {Piskunov, A. E.}, {Schilbach, E.}, {Röser, S.}, \& {Scholz, R.-D.}}]{Kharchenko2013}
{Kharchenko, N. V.}, {Piskunov, A. E.}, {Schilbach, E.}, {Röser, S.}, \& {Scholz, R.-D.} 2013, A\&A, 558, A53, \dodoi{10.1051/0004-6361/201322302}

\bibitem[{{Kipping}(2013)}]{Kipping2013}
{Kipping}, D.~M. 2013, \mnras, 435, 2152, \dodoi{10.1093/mnras/stt1435}

\bibitem[{{Koepferl} {et~al.}(2013){Koepferl}, {Ercolano}, {Dale}, {Teixeira}, {Ratzka}, \& {Spezzi}}]{Koepferl2013}
{Koepferl}, C.~M., {Ercolano}, B., {Dale}, J., {et~al.} 2013, \mnras, 428, 3327, \dodoi{10.1093/mnras/sts276}

\bibitem[{{Kounkel} {et~al.}(2020){Kounkel}, {Covey}, \& {Stassun}}]{Kounkel2020}
{Kounkel}, M., {Covey}, K., \& {Stassun}, K.~G. 2020, \aj, 160, 279, \dodoi{10.3847/1538-3881/abc0e6}

\bibitem[{{Kreidberg}(2015)}]{BATMAN}
{Kreidberg}, L. 2015, \pasp, 127, 1161, \dodoi{10.1086/683602}

\bibitem[{{Kunimoto} {et~al.}(2022){Kunimoto}, {Daylan}, {Guerrero}, {Fong}, {Bryson}, {Ricker}, {Fausnaugh}, {Huang}, {Sha}, {Shporer}, {Vanderburg}, {Vanderspek}, \& {Yu}}]{Kunimoto2022_QLPfaintstar}
{Kunimoto}, M., {Daylan}, T., {Guerrero}, N., {et~al.} 2022, \apjs, 259, 33, \dodoi{10.3847/1538-4365/ac5688}

\bibitem[{{Lee} {et~al.}(2014){Lee}, {Chiang}, \& {Ormel}}]{Lee2014}
{Lee}, E.~J., {Chiang}, E., \& {Ormel}, C.~W. 2014, \apj, 797, 95, \dodoi{10.1088/0004-637X/797/2/95}

\bibitem[{{Luque} \& {Pall{\'e}}(2022)}]{Luque2022}
{Luque}, R., \& {Pall{\'e}}, E. 2022, Science, 377, 1211, \dodoi{10.1126/science.abl7164}

\bibitem[{{Mann} {et~al.}(2013){Mann}, {Gaidos}, \& {Ansdell}}]{Mann2013c}
{Mann}, A.~W., {Gaidos}, E., \& {Ansdell}, M. 2013, \apj, 779, 188, \dodoi{10.1088/0004-637X/779/2/188}

\bibitem[{{Mann} {et~al.}(2016{\natexlab{a}}){Mann}, {Newton}, {Rizzuto}, {Irwin}, {Feiden}, {Gaidos}, {Mace}, {Kraus}, {James}, {Ansdell}, {Charbonneau}, {Covey}, {Ireland}, {Jaffe}, {Johnson}, {Kidder}, \& {Vanderburg}}]{Mann2016_k233}
{Mann}, A.~W., {Newton}, E.~R., {Rizzuto}, A.~C., {et~al.} 2016{\natexlab{a}}, \aj, 152, 61, \dodoi{10.3847/0004-6256/152/3/61}

\bibitem[{{Mann} {et~al.}(2016{\natexlab{b}}){Mann}, {Gaidos}, {Mace}, {Johnson}, {Bowler}, {LaCourse}, {Jacobs}, {Vanderburg}, {Kraus}, {Kaplan}, \& {Jaffe}}]{Mann2016a}
{Mann}, A.~W., {Gaidos}, E., {Mace}, G.~N., {et~al.} 2016{\natexlab{b}}, \apj, 818, 46, \dodoi{10.3847/0004-637X/818/1/46}

\bibitem[{{Mann} {et~al.}(2022){Mann}, {Wood}, {Schmidt}, {Barber}, {Owen}, {Tofflemire}, {Newton}, {Mamajek}, {Bush}, {Mace}, {Kraus}, {Thao}, {Vanderburg}, {Llama}, {Johns-Krull}, {Prato}, {Stahl}, {Tang}, {Fields}, {Collins}, {Collins}, {Gan}, {Jensen}, {Kamler}, {Schwarz}, {Furlan}, {Gnilka}, {Howell}, {Lester}, {Owens}, {Suarez}, {Mekarnia}, {Guillot}, {Abe}, {Triaud}, {Johnson}, {Milburn}, {Rizzuto}, {Quinn}, {Kerr}, {Ricker}, {Vanderspek}, {Latham}, {Seager}, {Winn}, {Jenkins}, {Guerrero}, {Shporer}, {Schlieder}, {McLean}, \& {Wohler}}]{THYMEVI_1227}
{Mann}, A.~W., {Wood}, M.~L., {Schmidt}, S.~P., {et~al.} 2022, \aj, 163, 156, \dodoi{10.3847/1538-3881/ac511d}

\bibitem[{{Marimbu} \& {Lee}(2024)}]{Marimbu2024}
{Marimbu}, K., \& {Lee}, E.~J. 2024, Research Notes of the American Astronomical Society, 8, 208, \dodoi{10.3847/2515-5172/ad7380}

\bibitem[{{Marley} {et~al.}(2021){Marley}, {Saumon}, {Visscher}, {Lupu}, {Freedman}, {Morley}, {Fortney}, {Seay}, {Smith}, {Teal}, \& {Wang}}]{Marley2021}
{Marley}, M.~S., {Saumon}, D., {Visscher}, C., {et~al.} 2021, \apj, 920, 85, \dodoi{10.3847/1538-4357/ac141d}

\bibitem[{{Masuda} \& {Winn}(2020)}]{Masuda2020}
{Masuda}, K., \& {Winn}, J.~N. 2020, \aj, 159, 81, \dodoi{10.3847/1538-3881/ab65be}

\bibitem[{{McCully} {et~al.}(2018){McCully}, {Volgenau}, {Harbeck}, {Lister}, {Saunders}, {Turner}, {Siiverd}, \& {Bowman}}]{McCully:2018}
{McCully}, C., {Volgenau}, N.~H., {Harbeck}, D.-R., {et~al.} 2018, in Society of Photo-Optical Instrumentation Engineers (SPIE) Conference Series, Vol. 10707, \procspie, 107070K, \dodoi{10.1117/12.2314340}

\bibitem[{{Mordasini} {et~al.}(2009){Mordasini}, {Alibert}, \& {Benz}}]{Mordasini2009}
{Mordasini}, C., {Alibert}, Y., \& {Benz}, W. 2009, \aap, 501, 1139, \dodoi{10.1051/0004-6361/200810301}

\bibitem[{{Nguyen, C. T.} {et~al.}(2022){Nguyen, C. T.}, {Costa, G.}, {Girardi, L.}, {Volpato, G.}, {Bressan, A.}, {Chen, Y.}, {Marigo, P.}, {Fu, X.}, \& {Goudfrooij, P.}}]{Nguyen2022_parsec20}
{Nguyen, C. T.}, {Costa, G.}, {Girardi, L.}, {et~al.} 2022, A\&A, 665, A126, \dodoi{10.1051/0004-6361/202244166}

\bibitem[{{Pang} {et~al.}(2022){Pang}, {Tang}, {Li}, {Yu}, {Wang}, {Li}, {Li}, {Wang}, {Wang}, {Zhang}, {Pasquato}, \& {Kouwenhoven}}]{Pang2022}
{Pang}, X., {Tang}, S.-Y., {Li}, Y., {et~al.} 2022, \apj, 931, 156, \dodoi{10.3847/1538-4357/ac674e}

\bibitem[{Paredes {et~al.}(2021)Paredes, Henry, Quinn, Gies, Hinojosa-Goñi, James, Jao, \& White}]{Paredes2021}
Paredes, L.~A., Henry, T.~J., Quinn, S.~N., {et~al.} 2021, The Astronomical Journal, 162, 176, \dodoi{10.3847/1538-3881/ac082a}

\bibitem[{{Parviainen} \& {Aigrain}(2015)}]{Parviainen2015}
{Parviainen}, H., \& {Aigrain}, S. 2015, \mnras, 453, 3821, \dodoi{10.1093/mnras/stv1857}

\bibitem[{{Qin} {et~al.}(2023){Qin}, {Zhong}, {Tang}, \& {Chen}}]{qin2023}
{Qin}, S., {Zhong}, J., {Tang}, T., \& {Chen}, L. 2023, \apjs, 265, 12, \dodoi{10.3847/1538-4365/acadd6}

\bibitem[{{Rayner} {et~al.}(2009){Rayner}, {Cushing}, \& {Vacca}}]{Rayner2009}
{Rayner}, J.~T., {Cushing}, M.~C., \& {Vacca}, W.~D. 2009, \apjs, 185, 289, \dodoi{10.1088/0067-0049/185/2/289}

\bibitem[{{Riello} {et~al.}(2021){Riello}, {De Angeli}, {Evans}, {Montegriffo}, {Carrasco}, {Busso}, {Palaversa}, {Burgess}, {Diener}, {Davidson}, {Rowell}, {Fabricius}, {Jordi}, {Bellazzini}, {Pancino}, {Harrison}, {Cacciari}, {van Leeuwen}, {Hambly}, {Hodgkin}, {Osborne}, {Altavilla}, {Barstow}, {Brown}, {Castellani}, {Cowell}, {De Luise}, {Gilmore}, {Giuffrida}, {Hidalgo}, {Holland}, {Marinoni}, {Pagani}, {Piersimoni}, {Pulone}, {Ragaini}, {Rainer}, {Richards}, {Sanna}, {Walton}, {Weiler}, \& {Yoldas}}]{Riello2021}
{Riello}, M., {De Angeli}, F., {Evans}, D.~W., {et~al.} 2021, \aap, 649, A3, \dodoi{10.1051/0004-6361/202039587}

\bibitem[{{Rizzuto} {et~al.}(2017){Rizzuto}, {Mann}, {Vanderburg}, {Kraus}, \& {Covey}}]{Rizzuto2017}
{Rizzuto}, A.~C., {Mann}, A.~W., {Vanderburg}, A., {Kraus}, A.~L., \& {Covey}, K.~R. 2017, \aj, 154, 224, \dodoi{10.3847/1538-3881/aa9070}

\bibitem[{{Rizzuto} {et~al.}(2020){Rizzuto}, {Newton}, {Mann}, {Tofflemire}, {Vanderburg}, {Kraus}, {Wood}, {Quinn}, {Zhou}, {Thao}, {Law}, {Ziegler}, \& {Brice{\~n}o}}]{THYMEII}
{Rizzuto}, A.~C., {Newton}, E.~R., {Mann}, A.~W., {et~al.} 2020, \aj, 160, 33, \dodoi{10.3847/1538-3881/ab94b7}

\bibitem[{{Rogers}(2025)}]{Rogers2025}
{Rogers}, J.~G. 2025, \mnras, 539, 2230, \dodoi{10.1093/mnras/staf628}

\bibitem[{{Rogers} {et~al.}(2024){Rogers}, {Owen}, \& {Schlichting}}]{Rogers2024}
{Rogers}, J.~G., {Owen}, J.~E., \& {Schlichting}, H.~E. 2024, \mnras, 529, 2716, \dodoi{10.1093/mnras/stae563}

\bibitem[{{Skrutskie} {et~al.}(2006){Skrutskie}, {Cutri}, {Stiening}, {Weinberg}, {Schneider}, {Carpenter}, {Beichman}, {Capps}, {Chester}, {Elias}, {Huchra}, {Liebert}, {Lonsdale}, {Monet}, {Price}, {Seitzer}, {Jarrett}, {Kirkpatrick}, {Gizis}, {Howard}, {Evans}, {Fowler}, {Fullmer}, {Hurt}, {Light}, {Kopan}, {Marsh}, {McCallon}, {Tam}, {Van Dyk}, \& {Wheelock}}]{Skrutskie2006}
{Skrutskie}, M.~F., {Cutri}, R.~M., {Stiening}, R., {et~al.} 2006, \aj, 131, 1163, \dodoi{10.1086/498708}

\bibitem[{{Tayar} {et~al.}(2022){Tayar}, {Claytor}, {Huber}, \& {van Saders}}]{Tayar2022}
{Tayar}, J., {Claytor}, Z.~R., {Huber}, D., \& {van Saders}, J. 2022, \apj, 927, 31, \dodoi{10.3847/1538-4357/ac4bbc}

\bibitem[{{TESS Team}(2021)}]{MAST_SPOC_LCs}
{TESS Team}. 2021, TESS "Fast" Light Curves - All Sectors,  STScI/MAST, \dodoi{10.17909/T9-ST5G-3177}

\bibitem[{{TESS Team}(2022)}]{MAST_FFI_LCs}
---. 2022, TESS Raw Full Frame Images: All Sectors,  STScI/MAST, \dodoi{10.17909/3Y7C-WA45}

\bibitem[{{Thao} {et~al.}(2023){Thao}, {Mann}, {Gao}, {Owens}, {Vanderburg}, {Newton}, {Tang}, {Fields}, {David}, {Irwin}, {Husser}, {Charbonneau}, \& {Ballard}}]{Thao2023}
{Thao}, P.~C., {Mann}, A.~W., {Gao}, P., {et~al.} 2023, \aj, 165, 23, \dodoi{10.3847/1538-3881/aca07a}

\bibitem[{{Thao} {et~al.}(2024){Thao}, {Mann}, {Feinstein}, {Gao}, {Thorngren}, {Rotman}, {Welbanks}, {Brown}, {Duvvuri}, {France}, {Longo}, {Sandoval}, {Schneider}, {Wilson}, {Youngblood}, {Vanderburg}, {Barber}, {Wood}, {Batalha}, {Kraus}, {Murray}, {Newton}, {Rizzuto}, {Tofflemire}, {Tsai}, {Bean}, {Berta-Thompson}, {Evans-Soma}, {Froning}, {Kempton}, {Miguel}, \& {Pineda}}]{Thao2024_featherweight}
{Thao}, P.~C., {Mann}, A.~W., {Feinstein}, A.~D., {et~al.} 2024, \aj, 168, 297, \dodoi{10.3847/1538-3881/ad81d7}

\bibitem[{Thao {et~al.}(2024)Thao, Mann, Barber, Kraus, Tofflemire, Bush, Wood, Collins, Vanderburg, Quinn, Zhou, Newton, Ziegler, Law, Barkaoui, Pozuelos, Timmermans, Gillon, Jehin, Schwarz, Gan, Shporer, Horne, Sefako, Suarez, Mekarnia, Guillot, Abe, Triaud, Radford, Murillo, Ricker, Winn, Jenkins, Bouma, Fausnaugh, Guerrero, \& Kunimoto}]{Thao2024_1224}
Thao, P.~C., Mann, A.~W., Barber, M.~G., {et~al.} 2024, \aj, 168, 41, \dodoi{10.3847/1538-3881/ad4993}

\bibitem[{{Tofflemire} {et~al.}(2021){Tofflemire}, {Rizzuto}, {Newton}, {Kraus}, {Mann}, {Vanderburg}, {Nelson}, {Hawkins}, {Wood}, {Zhou}, {Quinn}, {Howell}, {Collins}, {Schwarz}, {Stassun}, {Bouma}, {Essack}, {Osborn}, {Boyd}, {F{\H{u}}r{\'e}sz}, {Glidden}, {Twicken}, {Wohler}, {McLean}, {Ricker}, {Vanderspek}, {Latham}, {Seager}, {Winn}, \& {Jenkins}}]{THYMEV_FF}
{Tofflemire}, B.~M., {Rizzuto}, A.~C., {Newton}, E.~R., {et~al.} 2021, \aj, 161, 171, \dodoi{10.3847/1538-3881/abdf53}

\bibitem[{Tokovinin {et~al.}(2013)Tokovinin, Fischer, Bonati, Giguere, Moore, Schwab, Spronck, \& Szymkowiak}]{Tokovinin2013}
Tokovinin, A., Fischer, D.~A., Bonati, M., {et~al.} 2013, Publications of the Astronomical Society of the Pacific, 125, 1336, \dodoi{10.1086/674012}

\bibitem[{{Vach} {et~al.}(2024){Vach}, {Zhou}, {Huang}, {Rogers}, {Bouma}, {Douglas}, {Kunimoto}, {Mann}, {Barber}, {Quinn}, {Latham}, {Bieryla}, \& {Collins}}]{Vach2024_ocr}
{Vach}, S., {Zhou}, G., {Huang}, C.~X., {et~al.} 2024, \aj, 167, 210, \dodoi{10.3847/1538-3881/ad3108}

\bibitem[{{Vach} {et~al.}(2025){Vach}, {Zhou}, {Mann}, {Barber}, {Fairnington}, {Huang}, {Rogers}, {Bouma}, {Kr{\"u}ger}, {Wright}, {Niblett}, {Nelson}, {Quinn}, {Latham}, {Bieryla}, {Collins}, {Kunimoto}, {Watkins}, {Schwarz}, {Collins}, {Sefako}, {Horne}, {Howell}, {Clark}, {Littlefield}, {Christiansen}, {Essack}, \& {Winn}}]{Vach2025_tic887}
{Vach}, S., {Zhou}, G., {Mann}, A.~W., {et~al.} 2025, arXiv e-prints, arXiv:2502.00576, \dodoi{10.48550/arXiv.2502.00576}

\bibitem[{{Van Eylen} \& {Albrecht}(2015)}]{Van-Eylen2015}
{Van Eylen}, V., \& {Albrecht}, S. 2015, \apj, 808, 126, \dodoi{10.1088/0004-637X/808/2/126}

\bibitem[{{van Groeningen} {et~al.}(2023){van Groeningen}, {Castro-Ginard}, {Brown}, {Casamiquela}, \& {Jordi}}]{vanGroeningen2023}
{van Groeningen}, M.~G.~J., {Castro-Ginard}, A., {Brown}, A.~G.~A., {Casamiquela}, L., \& {Jordi}, C. 2023, \aap, 675, A68, \dodoi{10.1051/0004-6361/202345952}

\bibitem[{{Vanderburg} {et~al.}(2019){Vanderburg}, {Huang}, {Rodriguez}, {Becker}, {Ricker}, {Vanderspek}, {Latham}, {Seager}, {Winn}, {Jenkins}, {Addison}, {Bieryla}, {Brice{\~n}o}, {Bowler}, {Brown}, {Burke}, {Burt}, {Caldwell}, {Clark}, {Crossfield}, {Dittmann}, {Dynes}, {Fulton}, {Guerrero}, {Harbeck}, {Horner}, {Kane}, {Kielkopf}, {Kraus}, {Kreidberg}, {Law}, {Mann}, {Mengel}, {Morton}, {Okumura}, {Pearce}, {Plavchan}, {Quinn}, {Rabus}, {Rose}, {Rowden}, {Shporer}, {Siverd}, {Smith}, {Stassun}, {Tinney}, {Wittenmyer}, {Wright}, {Zhang}, {Zhou}, \& {Ziegler}}]{Vanderburg2019}
{Vanderburg}, A., {Huang}, C.~X., {Rodriguez}, J.~E., {et~al.} 2019, \apjl, 881, L19, \dodoi{10.3847/2041-8213/ab322d}

\bibitem[{{Venturini} {et~al.}(2016){Venturini}, {Alibert}, \& {Benz}}]{Venturini2016}
{Venturini}, J., {Alibert}, Y., \& {Benz}, W. 2016, \aap, 596, A90, \dodoi{10.1051/0004-6361/201628828}

\bibitem[{{Villaume} {et~al.}(2017){Villaume}, {Conroy}, {Johnson}, {Rayner}, {Mann}, \& {van Dokkum}}]{Villaume2017}
{Villaume}, A., {Conroy}, C., {Johnson}, B., {et~al.} 2017, \apjs, 230, 23, \dodoi{10.3847/1538-4365/aa72ed}

\bibitem[{{Weisserman} {et~al.}(2023){Weisserman}, {Becker}, \& {Vanderburg}}]{Weisserman2023}
{Weisserman}, D., {Becker}, J.~C., \& {Vanderburg}, A. 2023, \aj, 165, 89, \dodoi{10.3847/1538-3881/acac80}

\bibitem[{{Wirth} {et~al.}(2021){Wirth}, {Zhou}, {Quinn}, {Mann}, {Bouma}, {Latham}, {Teske}, {Wang}, {Shectman}, {Butler}, \& {Crane}}]{Wirth2021}
{Wirth}, C.~P., {Zhou}, G., {Quinn}, S.~N., {et~al.} 2021, \apjl, 917, L34, \dodoi{10.3847/2041-8213/ac13a9}

\bibitem[{{Wolf} {et~al.}(2018){Wolf}, {Onken}, {Luvaul}, {Schmidt}, {Bessell}, {Chang}, {Da Costa}, {Mackey}, {Martin-Jones}, {Murphy}, {Preston}, {Scalzo}, {Shao}, {Smillie}, {Tisserand}, {White}, \& {Yuan}}]{Skymapper1}
{Wolf}, C., {Onken}, C.~A., {Luvaul}, L.~C., {et~al.} 2018, \pasa, 35, e010, \dodoi{10.1017/pasa.2018.5}

\bibitem[{{Wolfgang} {et~al.}(2016){Wolfgang}, {Rogers}, \& {Ford}}]{Wolfgang2016}
{Wolfgang}, A., {Rogers}, L.~A., \& {Ford}, E.~B. 2016, \apj, 825, 19, \dodoi{10.3847/0004-637X/825/1/19}

\bibitem[{Wood {et~al.}(2021)Wood, Mann, \& Kraus}]{Wood2021_molusc}
Wood, M.~L., Mann, A.~W., \& Kraus, A.~L. 2021, The Astronomical Journal, 162, 128, \dodoi{10.3847/1538-3881/ac0ae9}

\bibitem[{{Wood} {et~al.}(2023){Wood}, {Mann}, {Barber}, {Bush}, {Kraus}, {Tofflemire}, {Vanderburg}, {Newton}, {Feiden}, {Zhou}, {Bouma}, {Quinn}, {Armstrong}, {Osborn}, {Adibekyan}, {Mena}, {Sousa}, {Gagn{\'e}}, {Fields}, {Milburn}, {Thao}, {Schmidt}, {Gnilka}, {Howell}, {Law}, {Ziegler}, {Brice{\~n}o}, {Ricker}, {Vanderspek}, {Latham}, {Seager}, {Winn}, {Jenkins}, {Schlieder}, {Osborn}, {Twicken}, {Ciardi}, \& {Huang}}]{Wood2023}
{Wood}, M.~L., {Mann}, A.~W., {Barber}, M.~G., {et~al.} 2023, \aj, 165, 85, \dodoi{10.3847/1538-3881/aca8fc}

\bibitem[{Wright \& Eastman(2014)}]{Wright2014_barycorrpy}
Wright, J.~T., \& Eastman, J.~D. 2014, Publications of the Astronomical Society of the Pacific, 126, 838, \dodoi{10.1086/678541}

\bibitem[{{Wu} \& {Lithwick}(2013)}]{Wu2013}
{Wu}, Y., \& {Lithwick}, Y. 2013, \apj, 772, 74, \dodoi{10.1088/0004-637X/772/1/74}

\bibitem[{{Zeng} {et~al.}(2019){Zeng}, {Jacobsen}, {Sasselov}, {Petaev}, {Vanderburg}, {Lopez-Morales}, {Perez-Mercader}, {Mattsson}, {Li}, {Heising}, {Bonomo}, {Damasso}, {Berger}, {Cao}, {Levi}, \& {Wordsworth}}]{Zeng2019}
{Zeng}, L., {Jacobsen}, S.~B., {Sasselov}, D.~D., {et~al.} 2019, Proceedings of the National Academy of Science, 116, 9723, \dodoi{10.1073/pnas.1812905116}

\bibitem[{{Zhou} {et~al.}(2020){Zhou}, {Winn}, {Newton}, {Quinn}, {Rodriguez}, {Mann}, {Rizzuto}, {Vanderburg}, {Huang}, {Latham}, {Teske}, {Wang}, {Shectman}, {Butler}, {Crane}, {Thompson}, {Henry}, {Paredes}, {Jao}, {James}, \& {Hinojosa}}]{Zhou2020}
{Zhou}, G., {Winn}, J.~N., {Newton}, E.~R., {et~al.} 2020, \apjl, 892, L21, \dodoi{10.3847/2041-8213/ab7d3c}

\bibitem[{{Ziegler} {et~al.}(2018){Ziegler}, {Law}, {Baranec}, {Morton}, {Riddle}, {De Lee}, {Huber}, {Mahadevan}, \& {Pepper}}]{Ziegler2018}
{Ziegler}, C., {Law}, N.~M., {Baranec}, C., {et~al.} 2018, \aj, 156, 259, \dodoi{10.3847/1538-3881/aad80a}

\end{thebibliography}

\clearpage

\startlongtable
\begin{deluxetable*}{cccccccccccccccccc}\label{tab:bigTable}
    \tablehead{
    \colhead{Gaia DR3} & \colhead{$\alpha$} & \colhead{$\delta$} & \colhead{Gmag} & \colhead{$B_p$-$R_p$} & \colhead{$V_{off}$} & \colhead{$RV_{exp}$\tablenotemark{a}} & \colhead{$RV$\tablenotemark{b}} & \colhead{$\sigma_{RV}$\tablenotemark{b}} & \colhead{$P_{rot}$} & \colhead{$\sigma_{Prot}$} & \colhead{Vela\tablenotemark{c}}\\
    \colhead{} & \colhead{(deg, J2016)} & \colhead{(deg, J2016)} & \colhead{(mag)} & \colhead{(mag)}  & \colhead{(km/s)} & \colhead{(km/s)} & \colhead{(km/s)} & \colhead{(km/s)} & \colhead{(days)} & \colhead{(days)} & \colhead{}  
    }
    \startdata
    5590921485126584832 &  110.1704 & -33.9223 & 12.77 &  0.86 &       0.0 &            19.1 &   19.1 &          2.06 &      2.41 &            0.04 &    Y \\
5604855080592114048 &  109.6015 & -31.8131 & 18.93 &  3.23 &      0.03 &           19.54 &    ... &           ... &     12.41 &            0.88 &    N \\
5592834467864569472 &  111.2801 & -31.9203 & 17.66 &  2.83 &      0.07 &           19.17 &    ... &           ... &     13.57 &            1.05 &    Y \\
5592608071548308864 &   111.936 & -32.3345 & 10.26 &  0.26 &      0.07 &           18.96 &  10.53 &          2.95 &      0.24 &           0.001 &    Y \\
5591821504113055104 &  114.1769 & -33.7603 & 19.56 &  2.94 &       0.1 &           18.25 &    ... &           ... &       ... &             ... &    N \\
5589656088982368768 &  111.0502 & -36.0676 & 15.47 &  2.33 &      0.11 &           18.56 &  -0.82 &         13.54 &     0.602 &           0.003 &    Y \\
5592540245416913664 &  111.9919 &  -32.825 & 17.90 &  2.78 &      0.13 &           18.88 &    ... &           ... &      20.0 &            2.27 &    N \\
5592861749497777664 &   110.585 & -32.0637 & 16.24 &  2.25 &      0.13 &            19.3 &    ... &           ... &     0.213 &           0.001 &    Y \\
\enddata
    \tablecomments{
    \tablenotetext{a}{Expected radial velocity for membership}
    \tablenotetext{b}{From \gaia DR3 \citep{GaiaCollaboration2023}}
    \tablenotetext{c}{Identified in Vela Population IV in \cite{CantatGaudin2019}.}
    }
    \end{deluxetable*}

\end{document}